          [2001/04/25 1.1 (PWD)]
\documentclass[letterpaper,twocolumn]{aa}

\def\spose#1{\hbox to 0pt{#1\hss}}
\def\lta{\mathrel{\spose{\lower 3pt\hbox{$\mathchar"218$}}
     \raise 2.0pt\hbox{$\mathchar"13C$}}}
\def\gta{\mathrel{\spose{\lower 3pt\hbox{$\mathchar"218$}}
     \raise 2.0pt\hbox{$\mathchar"13E$}}}

\def\p*{\phantom{*}}

\usepackage{graphicx,epsfig}

\usepackage{natbib}
\bibpunct[, ]{(}{)}{;}{a}{}{,}

\begin{document}

\authorrunning{Quarles et al.}

\titlerunning{The instability transition for the restricted 3-body problem. III.}

\title{The instability transition for the restricted 3-body problem \\
III.~The Lyapunov exponent criterion}

\author{
B. Quarles \inst{1},
J. Eberle \inst{1},
Z. E. Musielak \inst{1,2},
\and
M. Cuntz \inst{1,3}
}
\offprints{Z. E. Musielak}

\institute{
       Department of Physics, Science Hall, University of Texas at Arlington,
       Arlington, TX 76019-0059, USA; \\
       \email{[billyq,wjeberle,zmusielak,cuntz]@uta.edu}
\and
       Kiepenheuer-Institut f\"ur Sonnenphysik, Sch\"oneckstr. 6, 79104 Freiburg, Germany
\and
       Institut f\"ur Theoretische Astrophysik, Universit\"at Heidelberg,
       Albert \"Uberle Str. 2, 69120 Heidelberg, Germany
}

\date{Received <date> / Accepted <date>}

\abstract{}
{We establish a criterion for the stability of planetary orbits in stellar
binary systems by using Lyapunov exponents and power spectra for the special
case of the circular restricted 3-body problem (CR3BP).  The criterion 
augments our earlier results given in the two previous papers of this 
series where stability criteria have been developed based on the Jacobi
constant and the hodograph method.}
{The centerpiece of our method is the concept of Lyapunov exponents, which 
are incorporated into the analysis of orbital stability by integrating 
the Jacobian of the CR3BP and orthogonalizing the tangent vectors via a 
well-established algorithm originally developed by Wolf et al. 
The criterion for orbital stability based on the Lyapunov exponents is 
independently verified by using power spectra. The obtained results are 
compared to results presented in the two previous papers of this series.}     
{It is shown that the maximum Lyapunov exponent can be used as an indicator 
for chaotic behaviour of planetary orbits, which is consistent with previous
applications of this method, particularly studies for the Solar System.
The chaotic behaviour corresponds to either orbital stability or instability, 
and it depends solely on the mass ratio $\mu$ of the binary components and 
the initial distance ratio $\rho_0$ of the planet relative to the stellar 
separation distance.  Detailed case studies are presented for $\mu = 0.3$
and 0.5.  The stability limits are characterized based on the value
of the maximum Lyapunov exponent.  However, chaos theory as well as the
concept of Lyapunov time prevents us from predicting exactly when the planet
is ejected.  Our method is also able to indicate evidence of quasi-periodicity.}
{For different mass ratios of the stellar components, we are able to
characterize stability limits for the CR3BP based on the value of the maximum
Lyapunov exponent.  This theoretical result allows us to link the study
of planetary orbital stability to chaos theory noting that there is a large
array of literature on the properties and significance of Lyapunov exponents.
Although our results are given for the special case of the CR3BP, we expect
that it may be possible to augment the proposed Lyapunov exponent criterion
to studies of planets in generalized stellar binary systems, which is strongly
motivated by existing observational results as well as results expected from
ongoing and future planet search missions.}


\maketitle


\section{Introduction}

A classical problem within the realm of orbital stability studies for theoretical and
observed planets in stellar binary systems is the circular restricted 
3-body problem (CR3BP) \citep[e.g.,][]{sze67,roy05}.  The 
CR3BP describes the motion of a body of negligible\footnote{Negligible mass means 
that although the body's motion is influenced by the gravity of the two massive 
primaries, its mass is too low to affect the motions of the primaries.} mass 
moving in the gravitational field of the two massive primaries considered here
to be two stars.  The stars move in circular orbits about the center of mass 
and their motion is not influenced by the third body, the planet.  Furthermore, 
the initial velocity of the planet is assumed in the same direction as the 
orbital velocity of its host star, which is usually the more massive of the two 
stars.

The study of planetary orbital stability is timely for various astronomical
reasons.  First, although most planets are found in wide binaries,
various cases of planets in binaries with separation distances of less than
30~AU have also been identified \citep[e.g.,][and references therein]{pat02,egg04,egg10}. 
Second, many more cases of planets in stellar binary systems are expected to
be discovered noting that binary (and possibly multiple) stellar systems
occur in high frequency in the local Galactic neighbourhood
\citep{duq91,lad06,rag06}.  Moreover, orbital stability studies of
planets around stars, including binary systems, are highly significant in
consideration of ongoing and future planet search missions
\citep[e.g.,][]{cat06,coc09}.

The CR3BP has been studied in detail by \cite{dvo84}, \cite{dvo86}, \cite{rab88}, 
\cite{smi93}, \cite{pil02}, and \cite{mar07}.  It has also been the focus 
of the previous papers in this series.  In Paper~I, \cite{ebe08} obtained the 
planetary stability limits using a criterion based on Jacobi's integral and
Jacobi's constant.  The method, also related to the concept of Hill stability 
\citep{roy05}, showed that orbital stability can be guaranteed only if the 
initial position of the planet lies within a well-defined limit determined by
the mass ratio of the stellar components.  Additionally, the stability criterion
was found to be also related to the borders of the ``zero velocity contour" (ZVC)
and its topology assessed by using a synodic coordinate system regarding the
two stellar components.

In Paper~II, \cite{ebe10} followed another theoretical concept based on the
hodograph eccentricity criterion.  This method relies on an 
approach given by differential geometry that analyzes the curvature of the 
planetary orbit in the synodic coordinate system.  The centerpiece of this 
method is the evaluation of the effective time-dependent eccentricity of the 
orbit based on the hodograph in rotating coordinates as well as the calculation
of the mean and median values of the eccentricity distribution.  This approach 
has been successful in mapping quasi-periodicity and multi-periodicity 
for planets in binary systems.  It has also been tested by comparing 
its theoretical predictions with work by \cite{hol99} and \cite{mus05} 
in regard to the extent of the region of orbital stability in binary 
systems of different mass ratios.

Previously, the work by \cite{ebe10} dealt with the assessment of short-time
orbital stability, encompassing time scales of 10$^3$~yrs or less.  One of the 
findings was the identification of a quasi-periodic region in the stellar mass 
and distance ($\mu$, $\rho_{0}$) parameter space (see Sect. 2.1).
Due to the relatively short 
time scales considered in the previous study, there is a strong motivation to 
revisit the onset of orbital instability using longer time scales and different 
types of methods.
The focus of this paper is the analysis of orbital stability by Lyapunov 
exponents, which are among the most commonly used numerical tools for investigating
chaotic behaviour of different dynamical systems \citep[e.g.,][]{hil94}.  The 
exponents have already repeatedly been used in orbital mechanics studies of
the Solar System \citep[e.g.,][]{lis99,mur01}.  For example, \cite{lis99}
discussed the long-term stability of the eight Solar System planets, while also
considering previous studies.  He concluded that the Solar System is most likely
astronomically stable, in the view of the limited life time of the Sun; however,
the orbits of Pluto and of many asteroids may become unstable soon after
the Sun becomes a white dwarf.

The numerical procedure of computing the Lyapunov exponents has been developed by   
\cite{wol85} based on previous work by \cite{ben80}.  Hence, the main objective
of the present paper is to establish the Lyapunov
exponent criterion and verify it by performing the power spectra analysis  
\citep[e.g.,][]{hil94} as well as by comparing the obtained results to those
presented in Paper I and II.  Our newly established criterion is then used to
investigate the stability of planetary orbits in stellar binary systems (approximated 
here as the CR3BP) of different mass and distance ratios.  The methodology of 
our study of orbital stability based on the Lyapunov exponent criterion is presented 
and discussed in \S2.  Detailed model simulations are described in \S3, and our
conclusions are given in \S4.

\section{Theoretical approach}

\subsection{Basic equations}

In the following, we consider the so-called coplanar
CR3BP \citep[e.g.,][]{sze67,roy05}, which we will define
as follows.  Two stars are in circular motion about their common center of 
mass and their masses are much larger than the mass of the planet.  In our 
case, the planetary mass is assumed to be $1 \times 10^{-6}$ of the mass 
of the star it orbits; also note that the planetary motion is constrained 
to the orbital plane of the two stars. Moreover, it is assumed that the 
initial velocity of the planet is in the same direction as the orbital 
velocity of its host star, which is the more massive of the two stars, 
and that the starting position of the planet is to the right of the 
primary star (3 o'clock position), along the line joining the binary 
components.

Following the conventions described by \cite{ebe08}, we write the 
equations of motion in terms of the parameters $\mu$ and $\rho_0$ with
$\mu$ and $\alpha = 1 - \mu$ being related to the ratio of the
two stellar masses $m_1$ and $m_2$ (see below).  Moreover, $R_0$
denotes the initial distance of the planet from its host star, the
more massive of the two stars with mass $m_1$, whereas $D$ denotes
the distance between the two stars, allowing us to define $\rho_0$.
In addition, we use a
rotating reference frame, which also gives rise to Coriolis and
centrifugal forces.  The equations of motion are given as
\begin{eqnarray}
\dot{x} & = & u \\
\dot{y} & = & v \\
\dot{z} & = & w \\
\dot{u} & = & 2v + x -\alpha {x - \mu \over r_{1}^3} -\mu {x + \alpha 
\over r_{2}^3} \\
\dot{v} & = & -2u + y -\alpha {y \over r_{1}^3} -\mu {y \over r_{2}^3}               \\
\dot{w} & = & -\alpha {z \over r_{1}^3} -\mu {z \over r_{2}^3}
\end{eqnarray}
where
\begin{eqnarray}
\mu     & = & {m_{2} \over m_{1} + m_{2}} \\
\alpha  & = & 1 - \mu \\
r_{1}^2 & = & \left(x - \mu\right)^2 + y^2 + z^2 \\
r_{2}^2 & = & \left(x + \alpha\right)^2 + y^2 + z^2\\
\rho_0 & = & {R_0 \over D}
\end{eqnarray}

The variables in the above equations describe the position of the planet,
which in essence constitutes a test particle.  Its position is defined in
Cartesian coordinates $\left\{x,\;y,\;z\right\}$.  
We denote the time derivative or velocity of a coordinate using the dot 
notation $\left\{\dot{x}={\rm dx \over dt}\right\}$.  We also represent
the set of second order differential equations, the equations
of motion, by a set of first order differential equations.  The velocity is
defined by the coordinates $\left\{u,\;v,\;w\right\}$ whose time derivatives
are the accelerations.  By defining the mass ratio and using normalized
coordinates, we can define the distances $\left\{r_{1},\;r_{2}\right\}$ with
reference to the location of the stars in the rotating coordinate system.

\begin{figure}
\centering
\epsfig{file=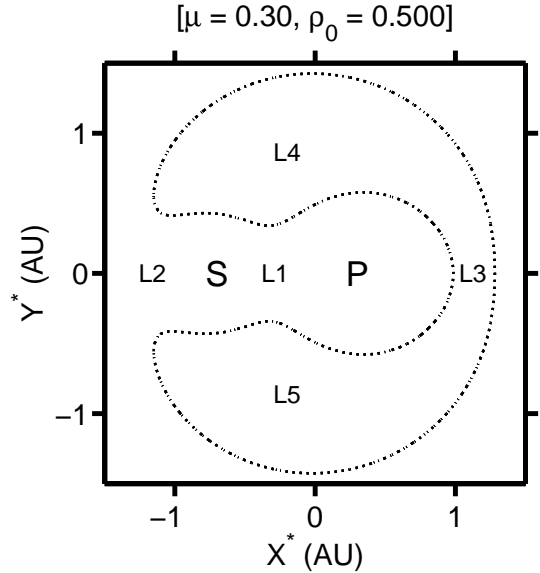,width=0.85\linewidth}
\caption{Locations of the Lagrange points L1, L2, L3,
L4, and L5 as used in the CR3BP.  The label P denotes the
primary star and the label S denotes the secondary star.}
\label{fig:convention}

\end{figure}

We enumerate the co-linear Lagrange points in the synodic frame by the order 
of which the ZVC opens.  The point between the stars opens first; therefore,
we denote it as L1.  The point to the left of the star that does not host the
planet opens second; thus, we denote it L2.  The point to the right of the star
hosting the planet opens third and it is denoted L3.  The two Trojan 
Lagrange points which are of lesser importance to our study are
L4 and L5 (see Fig.~\ref{fig:convention}).

\subsection{Lyapunov exponents}

A fundamental difference between stable and unstable planetary orbits is that
two nearby trajectories in phase space will diverge as a power law (usually
linear) for the former and exponentially for the latter. The parameter that 
is used to measure this rate of divergence is called Lyapunov exponent as it 
was originally introduced by \cite{lya07}; see also \cite{bak90}.  A 
dynamical system with {\it n} degrees of freedom is represented in {\it 2n} 
phase space; thus, to fully determine the stability of the system all {\it 2n} 
Lyapunov exponents must be calculated.  The Lyapunov exponents are the most 
commonly used tools to determine the onset of chaos and chaotic behaviour of 
both dissipative \citep[e.g.,][and references therein]{mus09} and non-dissipative
systems of orbital mechanics \citep[e.g.,][and references therein]{lis99}.  The
positive Lyapunov exponents measure the rate of divergence of neighbouring orbits,
whereas negative exponents measure the convergence rates between stable manifolds.
For dissipative dynamical systems the sum of all Lyapunov exponents is less than $0$
\citep[e.g.,][]{mus09}; however, for Hamiltonian (non-dissipative) systems the sum
is equal to $0$ \citep[e.g.,][]{hil94}.
  
Specific applications of the Lyapunov exponents to the circular restricted 3-body 
problem were discussed by many authors, including \cite{gon81}, \cite{jef83}, 
\cite{lec92}, \cite{mil92}, \cite{smi93}, \cite{mur01}, and others.
Some of these authors also considered the 
so-called Lyapunov time, which measures the $e$-folding time for the divergence
of nearby trajectories. It should be noted that the Lyapunov time is not 
well-defined for cases when close encounters between a planet and one of
the stars occur or when the planet is ejected from the system \citep{lis99}.
We shall treat such cases with special caution in this paper.      

The previously obtained results clearly show that the Lyapunov exponents can be 
calculated for the case of the CR3BP considered in this paper, for which we have 
$\textit{2n}=6$.  This requires a state vector for the system containing $6$ 
elements.  Details about the nature of the state vector are given in Appendix A.  
From the equations of motion, the Jacobian \textbf{J} can be determined, which
will be the foundation of how the Lyapunov exponents will be determined.  In 
addition, the nature of this Jacobian will elude to certain properties of the 
Lyapunov exponents.

For Hamiltonian systems the trace of the Jacobian should equal zero, $Tr\;\textbf{J}=0$.  
This requires that the CR3BP should have either all zero diagonal elements or an 
even amount of 3 positive/negative diagonal elements.  As a result this forces the 
sum of the Lyapunov exponents to be also zero.  Then we should expect some symmetry 
in the spectrum of the Lyapunov exponents.  Since this is indeed the case, the Lyapunov 
exponent spectra shown in this paper will only present the positive Lyapunov exponents 
and omit the negative Lyapunov exponents as they do not give any additional information 
about the system.  It is the convention that positive Lyapunov exponents indicate that 
both dissipative \citep[e.g.,][]{hil94} and non-dissipative \citep[e.g.,][]{ozo90}
systems are chaotic.  Here the end value of the Lyapunov exponents will be used to make 
the distinction between chaotic and non-chaotic orbits.

It is well known that the largest Lyapunov exponent is sufficient to determine the 
magnitude of the chaos in a system \citep[e.g.,][]{wol85}.  Therefore, this study
will consider only the effect of the maximum Lyapunov exponent because it will show
the greatest effect of chaos on the system.  This motivates us to also examine how
large the maximum Lyapunov exponent can be before introducing enough chaos to
affect the stability of the CR3BP.

\begin{table*}
\caption{Errors in the Jacobi constant (JCE) for models of $\mu = 0.3$
at different time intervals.}
\centering
\vspace{0.05in}
\vspace{0.05in}
\begin{tabular}{l c c c c }
\hline
\hline
\noalign{\vspace{0.03in}}
$\rho_0$ & $\rm JCE({\tau \over 4})$ & $\rm JCE({\tau \over 2})$ 
& $\rm JCE({3\tau \over 4})$ & $\rm JCE({\tau})$ \\
\noalign{\vspace{0.03in}}
\hline
\noalign{\vspace{0.03in}}
 0.355  & 1.0740E$-10$\p* &  2.4747E$-10$\p* &  1.2403E$-10$\p* & 1.8824E$-10$\p*        \\
 0.474  & 1.4498E$-10$\p* &  9.7917E$-11$\p* &  3.7723E$-11$\p* & 1.0067E$-11$\p*        \\
 0.595  & 8.9819E$-13$\p* &  5.2923E$-13$\p* &  5.2001E$-13$\p* & 7.1001E$-06$\p*        \\  
\noalign{\vspace{0.05in}} \hline
\end{tabular}
\vspace{0.05in}
\begin{list}{}{}
\item[]Note: $\tau$ denotes the runtime of the simulation.  We show the stable cases
of $\rho_0 = 0.355$ and 0.474 with $\tau = 10^5$ binary orbits as well as the unstable
case of $\rho_0 = 0.595$ with $\tau = 150.64$ binary orbits.
\end{list}
\label{table1}
\end{table*}

\begin{table*} 
\caption{Maximum Lyapunov exponent study for the models of $\mu = 0.3$ at different 
time intervals.}
\centering
\vspace{0.05in}
\vspace{0.05in}
\begin{tabular}{l c c c c c}
\hline
\hline
\noalign{\vspace{0.03in}}
$\rho_0$ & $\rm MLE\left({\rm 10^2}\right)$ & $\rm MLE\left({\rm 10^3}\right)$ 
& $\rm MLE\left({\rm 10^4}\right)$ & $e_{\rm median}$ & \rm ZVC  \\
\noalign{\vspace{0.03in}}
\hline
\noalign{\vspace{0.03in}}
 0.20  &  1.1099E$-1$\p*  &  1.1493E$-2$\p*  &  1.1446E$-3$\p* & 0.027 & ...        \\
 0.30  &  1.0322E$-1$\p*  &  1.0103E$-2$\p*  &  1.0691E$-3$\p* & 0.17  & ...        \\
 0.40  &  9.8084E$-2$\p*  &  1.0050E$-2$\p*  &  6.8986E$-4$\p* & 0.58  & L1         \\
 0.474 &  7.7766E$-2$\p*  &  9.4232E$-3$\p*  &  9.2624E$-4$\p* & 0.75  & L1, L2     \\
 0.50  &  8.5390E$-2$\p*  &  9.2048E$-3$\p*  &  7.9830E$-4$\p* & 0.85  & L1, L2     \\
 0.595 &  1.5793E$-1$\p*  &  1.4175E$-1$*    &   ...           & 1.14  & L1, L2, L3 \\
 0.60  &  8.9533E$-1$*    &   ...            &   ...           & 1.10  & L1, L2, L3 \\
\noalign{\vspace{0.05in}}
\hline
\end{tabular}
\vspace{0.05in}
\begin{list}{}{}
\item[]Note: Elements without data indicate simulations that ended due to the energy criterion,
thus representing planetary catastrophes.  Elements with an asterisk (*) indicate that the
simulation ended before the allotted time.  Also, $e_{\rm median}$ represents the median of
the eccentricity distribution obtained for the curvature of the planetary orbits in the
synodic coordinate system (see Paper~II).  The last column indicates the Lagrange point(s)
where the zero-velocity contour (ZVC) is open (see Paper~I).
\end{list}
\label{table2}
\end{table*}

\section{Results and discussion}

We performed simulations for stellar mass ratios from $\mu = 0.0$ to 0.5 in 
increments of 0.01.  A Yoshida sixth order symplectic integration scheme was used
\citep[e.g.,][]{yos90}.  As a measure of the precision of the integration scheme, we 
note that a time step of $\epsilon = 10^{-4}$~yrs is used for the individual cases.  
However, for producing plots of the entire parameter space we adopt a time step of
$\epsilon = 10^{-3}$~yrs as the smaller time step did not noticeably enhance the
quality of the plots.  The order of error in 
energy for each time step was $10^{-14}$ and $10^{-10}$, respectively.  We performed 
simulations for different time limits, which range from 10 to $10^5$~yrs in increments 
of powers of 10.  We also performed case studies with time limits of $10^6$ and $10^7$ yrs,  
although we do not expect significant changes to occur at those longer time scales.  By using 
different time scales we can estimate when certain phenomena occur and ascertain 
how they will affect other regions over longer periods of time.

\begin{figure*} [h]
\centering
\begin{tabular} {ccc}
\epsfig{file=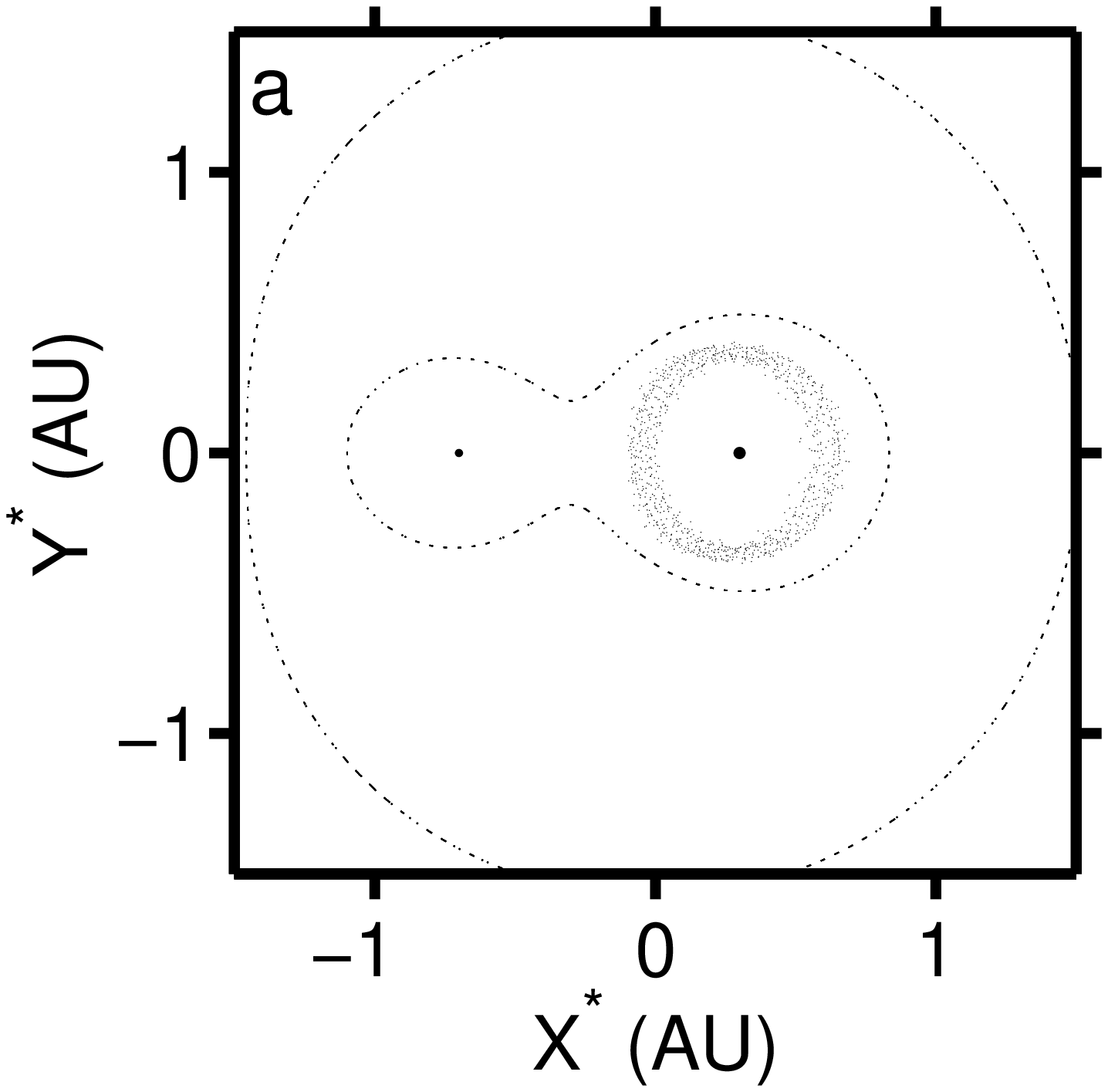,width=0.3\linewidth,height=0.3\linewidth}&
\epsfig{file=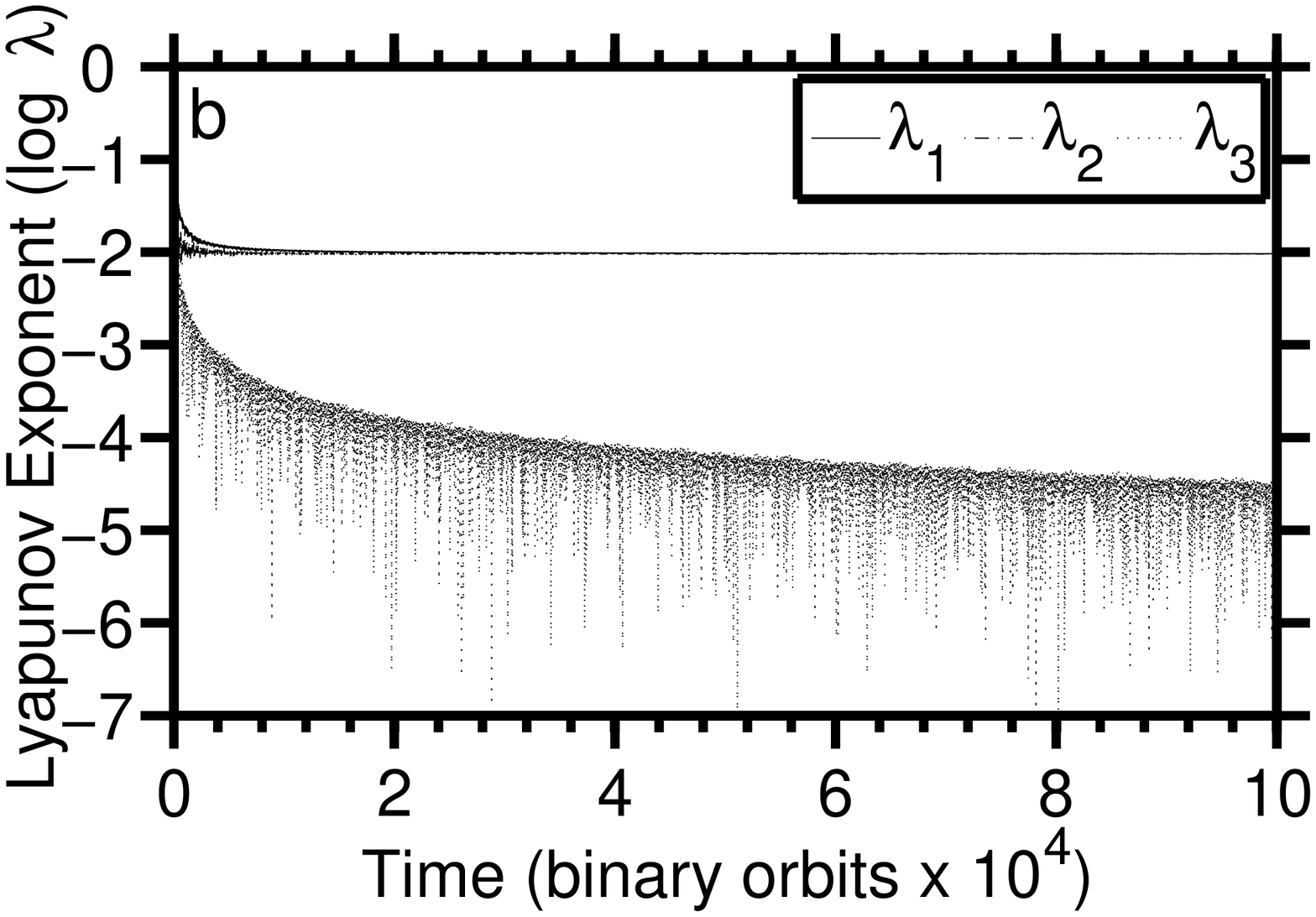,width=0.3\linewidth,height=0.3\linewidth}&
\epsfig{file=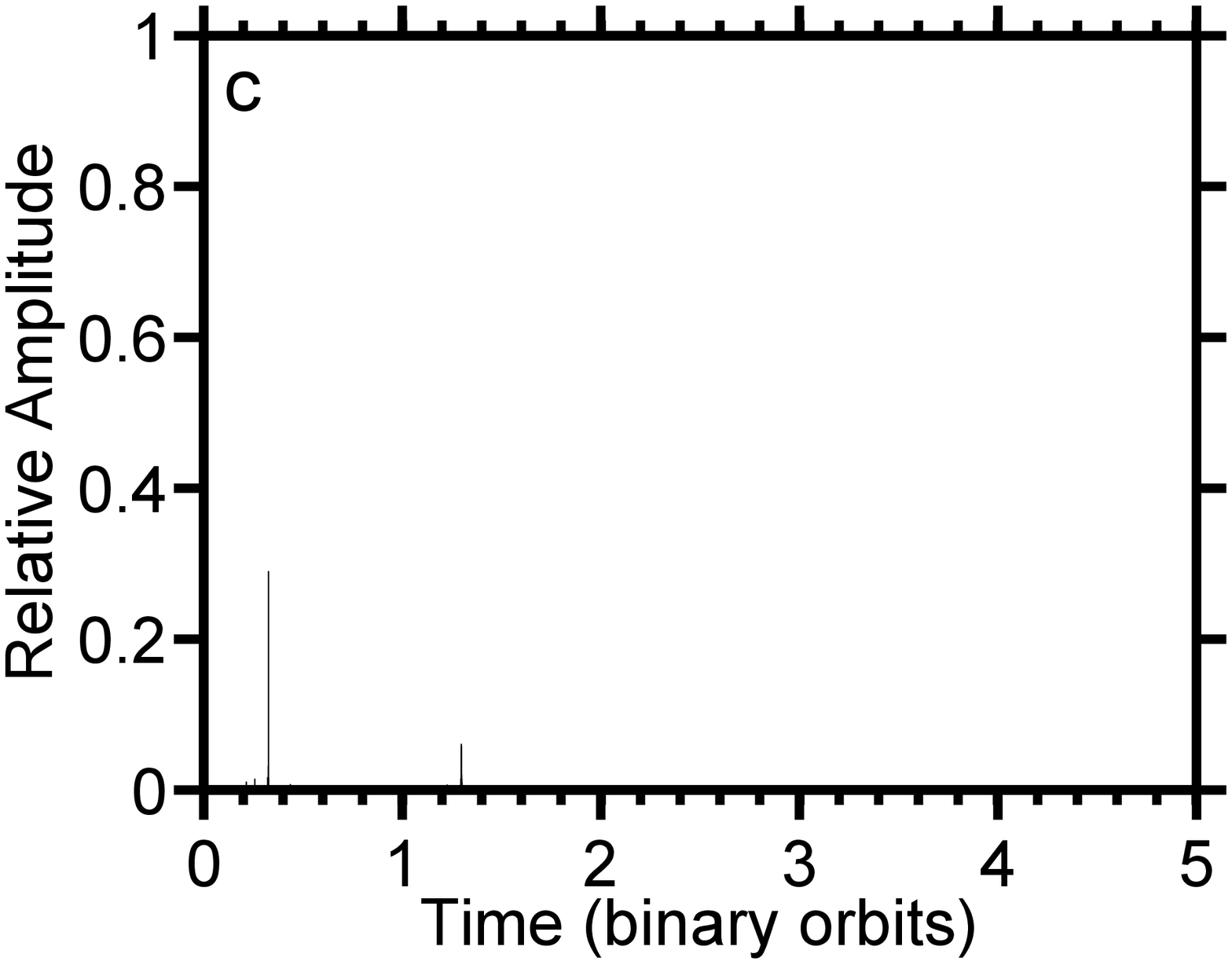,width=0.3\linewidth,height=0.3\linewidth}\\
\end{tabular}
\caption{Case study for the initial planetary distance ratio
$\rho_0 = 0.355$ with the planetary orbit in the synodic coordinate system, Lyapunov spectrum, 
and power spectrum.  The stellar mass ratio is $\mu = 0.3$.}
\label{fig:mu.3rho.355}
\end{figure*}

\begin{figure*} [h]
\centering
\begin{tabular} {ccc}
\epsfig{file=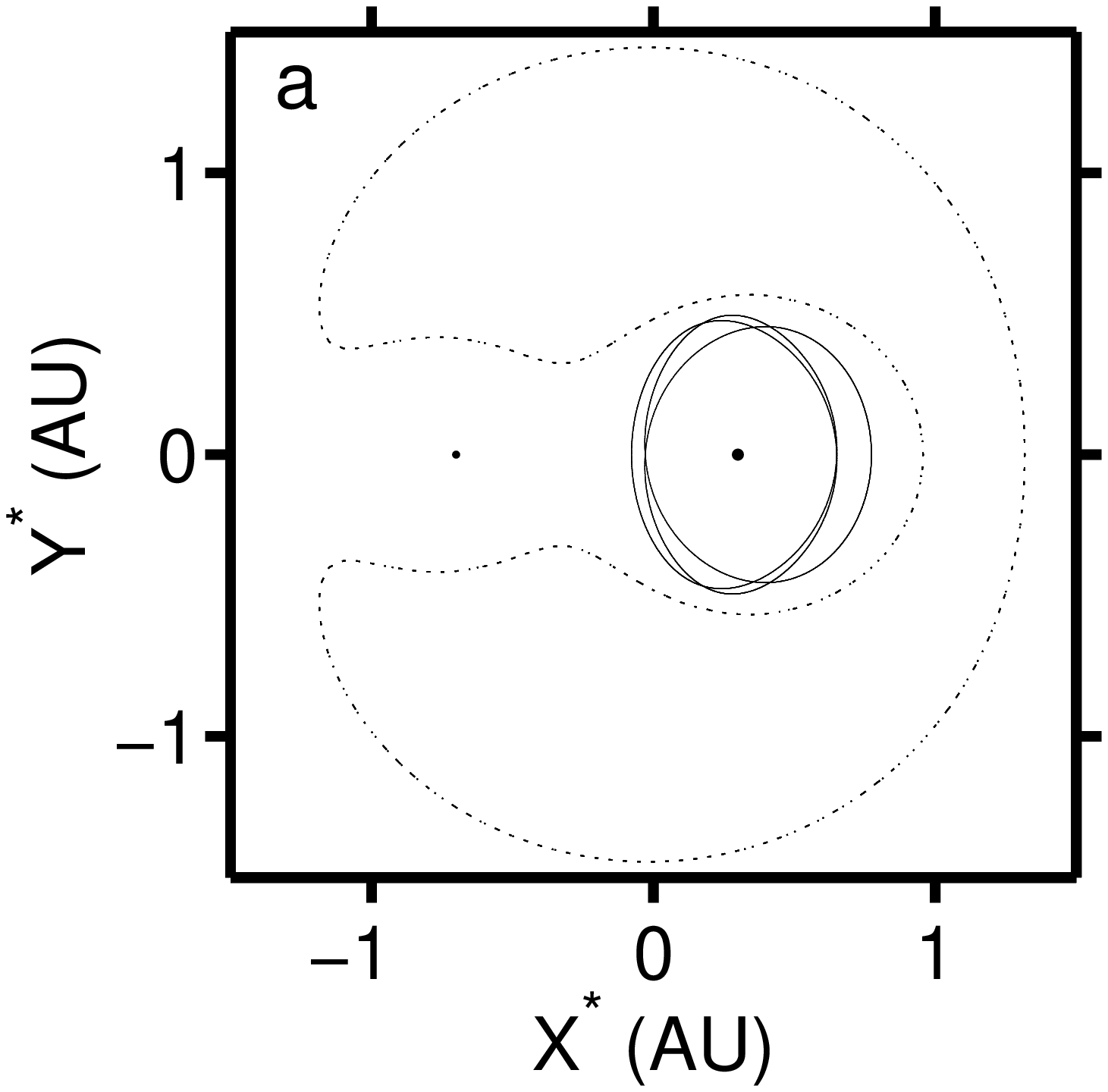,width=0.3\linewidth,height=0.3\linewidth}&
\epsfig{file=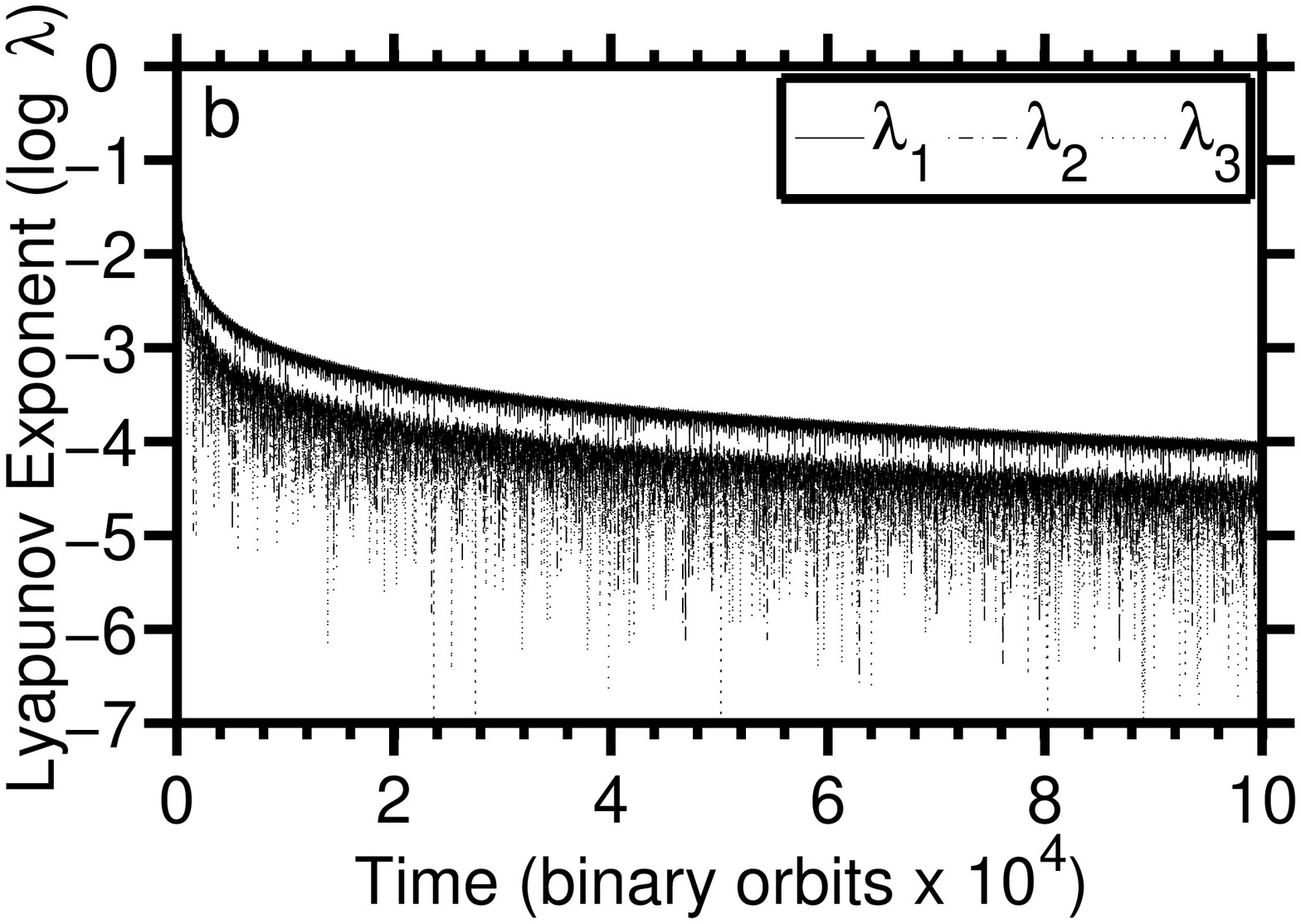,width=0.3\linewidth,height=0.3\linewidth}&
\epsfig{file=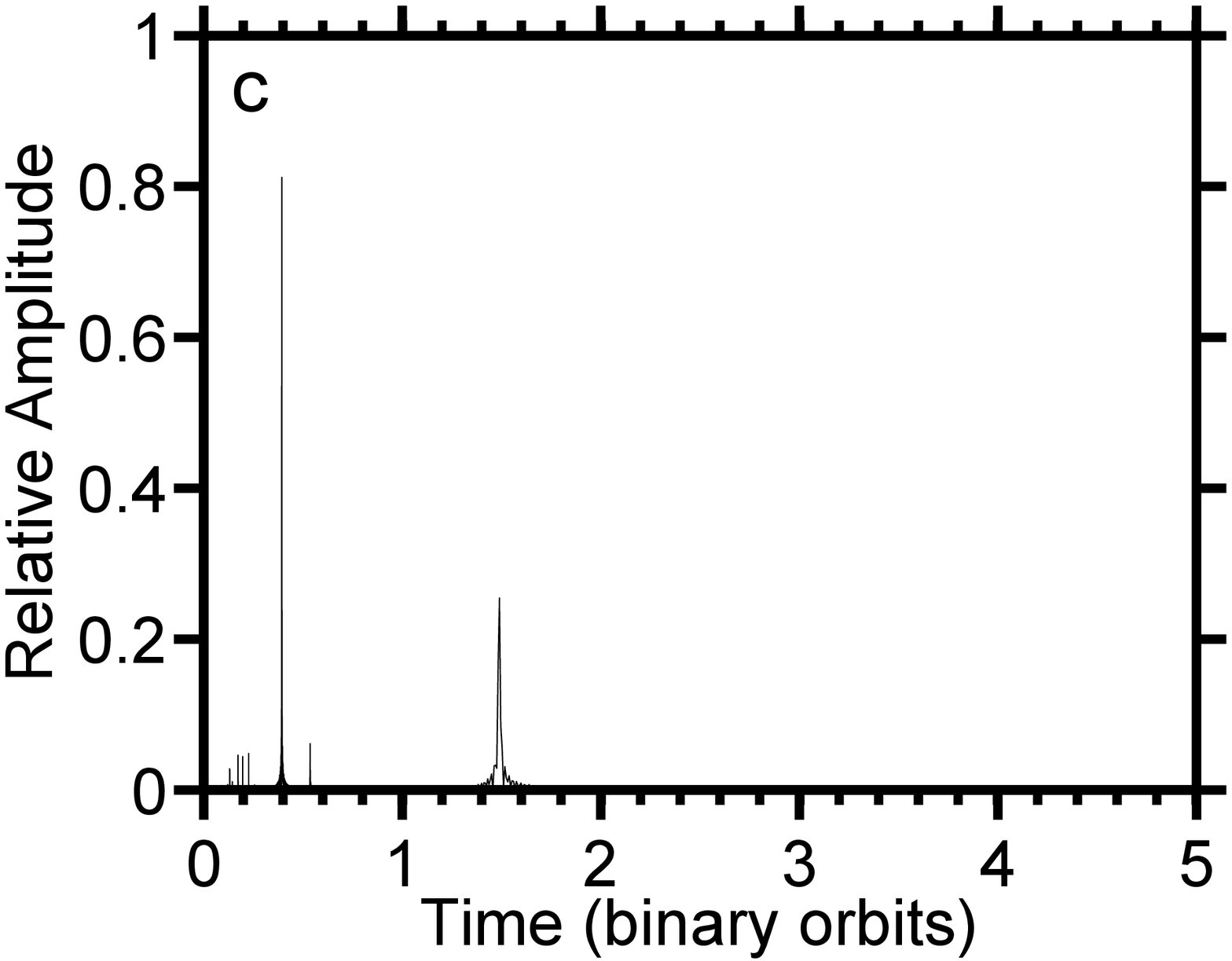,width=0.3\linewidth,height=0.3\linewidth}\\
\end{tabular}
\caption{Case study for the initial planetary distance ratio
$\rho_0 = 0.474$ with the planetary orbit in the synodic coordinate system, Lyapunov spectrum, 
and power spectrum.  The stellar mass ratio is $\mu = 0.3$.}
\label{fig:mu.3rho.474}
\end{figure*}

\begin{figure*} [h]
\centering
\begin{tabular} {ccc}
\epsfig{file=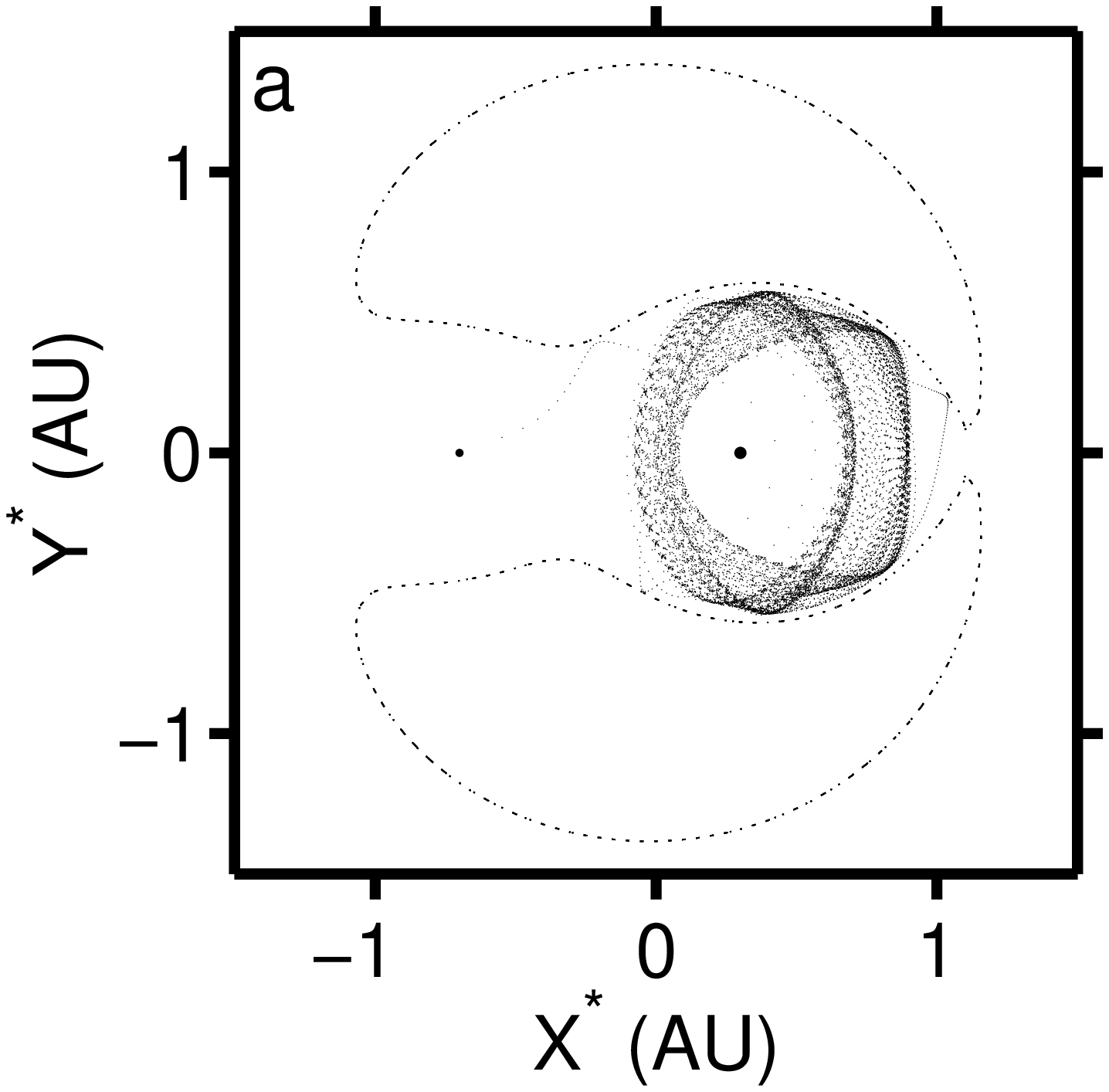,width=0.3\linewidth,height=0.3\linewidth}&
\epsfig{file=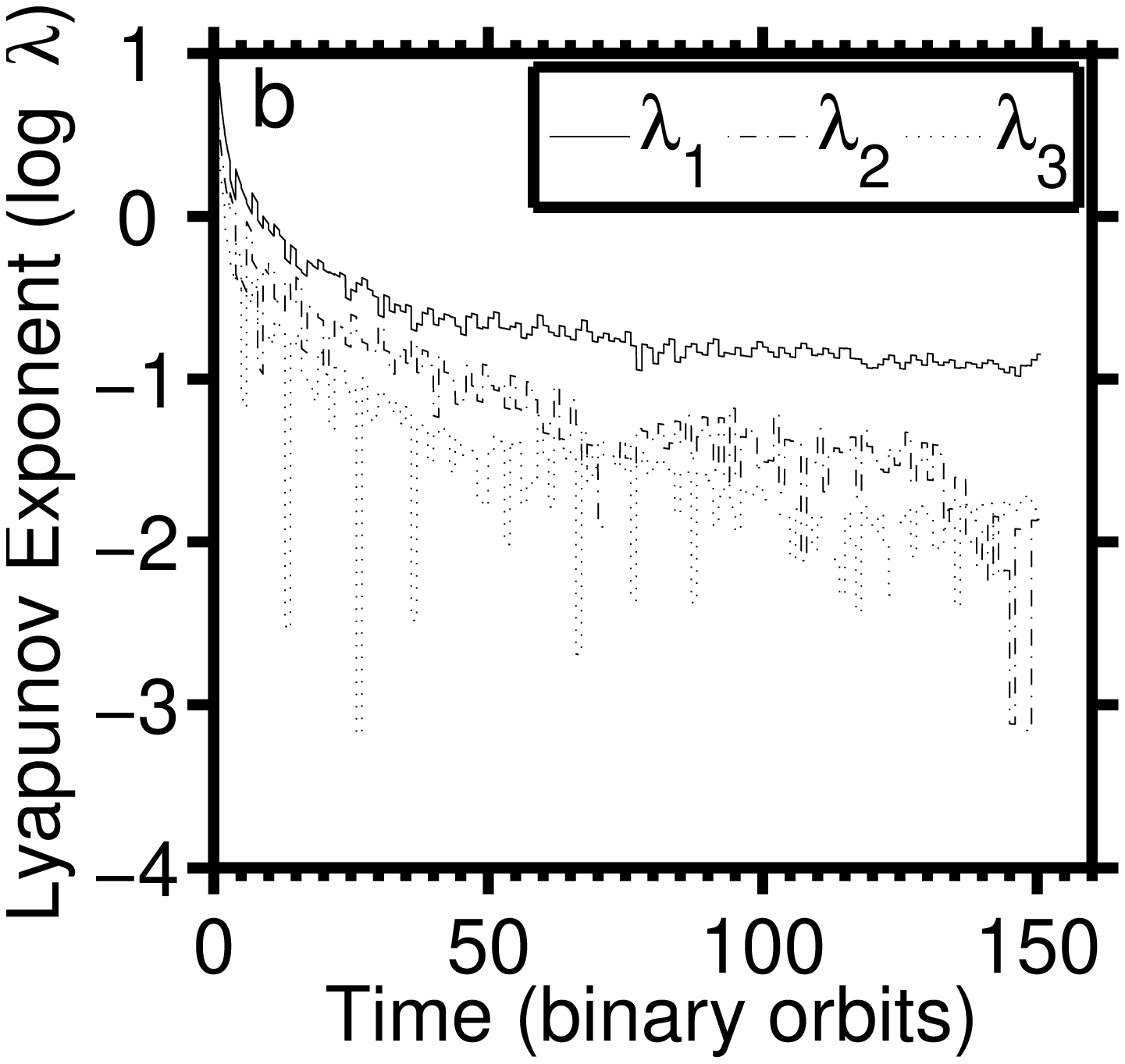,width=0.3\linewidth,height=0.3\linewidth}&
\epsfig{file=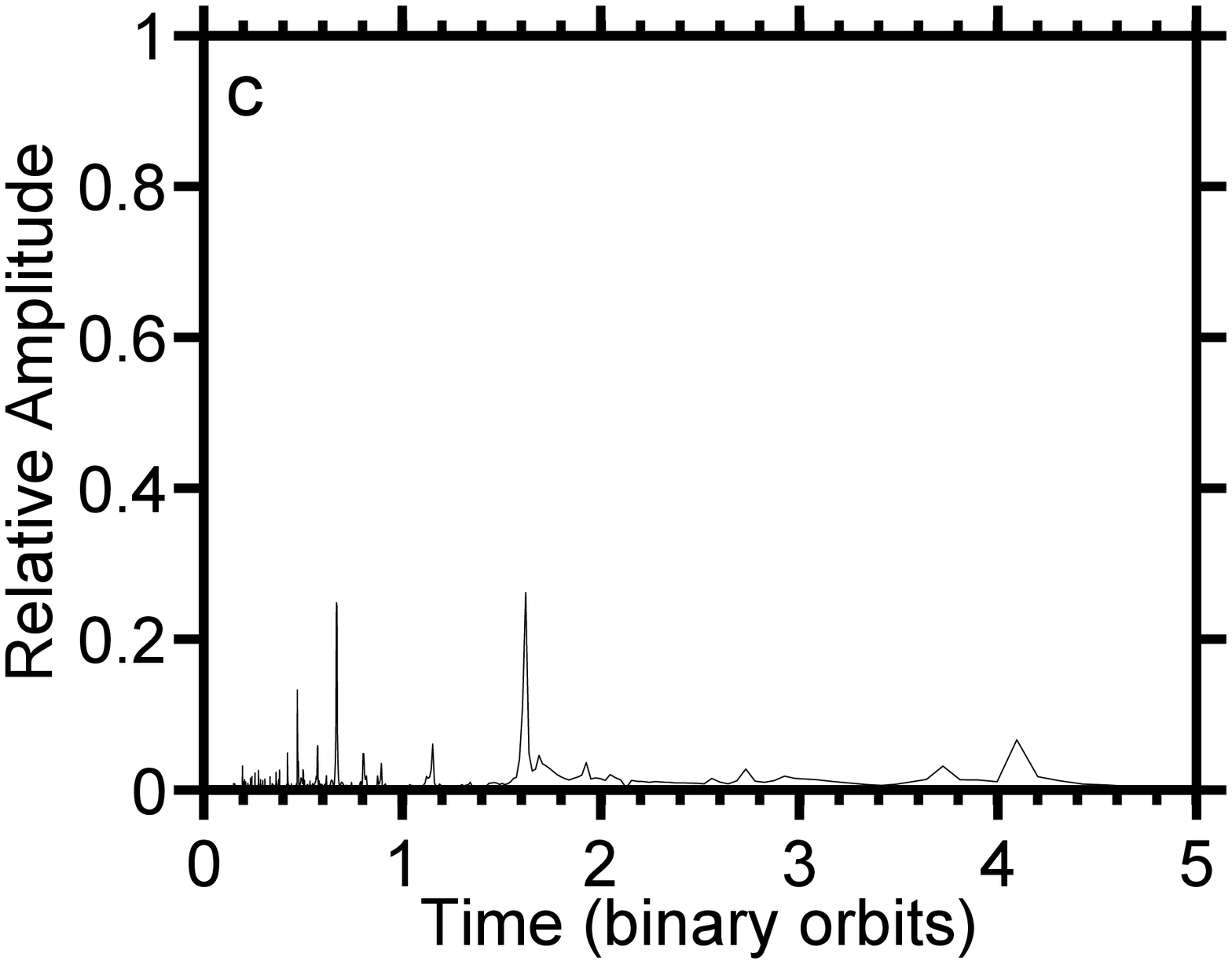,width=0.3\linewidth,height=0.3\linewidth}\\
\end{tabular}
\caption{Case study for the initial planetary distance ratio
$\rho_0 = 0.595$ with the planetary orbit in the synodic coordinate system, Lyapunov spectrum, 
and power spectrum.  The stellar mass ratio is $\mu = 0.3$.}
\label{fig:mu.3rho.595}
\end{figure*}

\begin{figure*} [h]
\centering
\begin{tabular} {ccc}
\epsfig{file=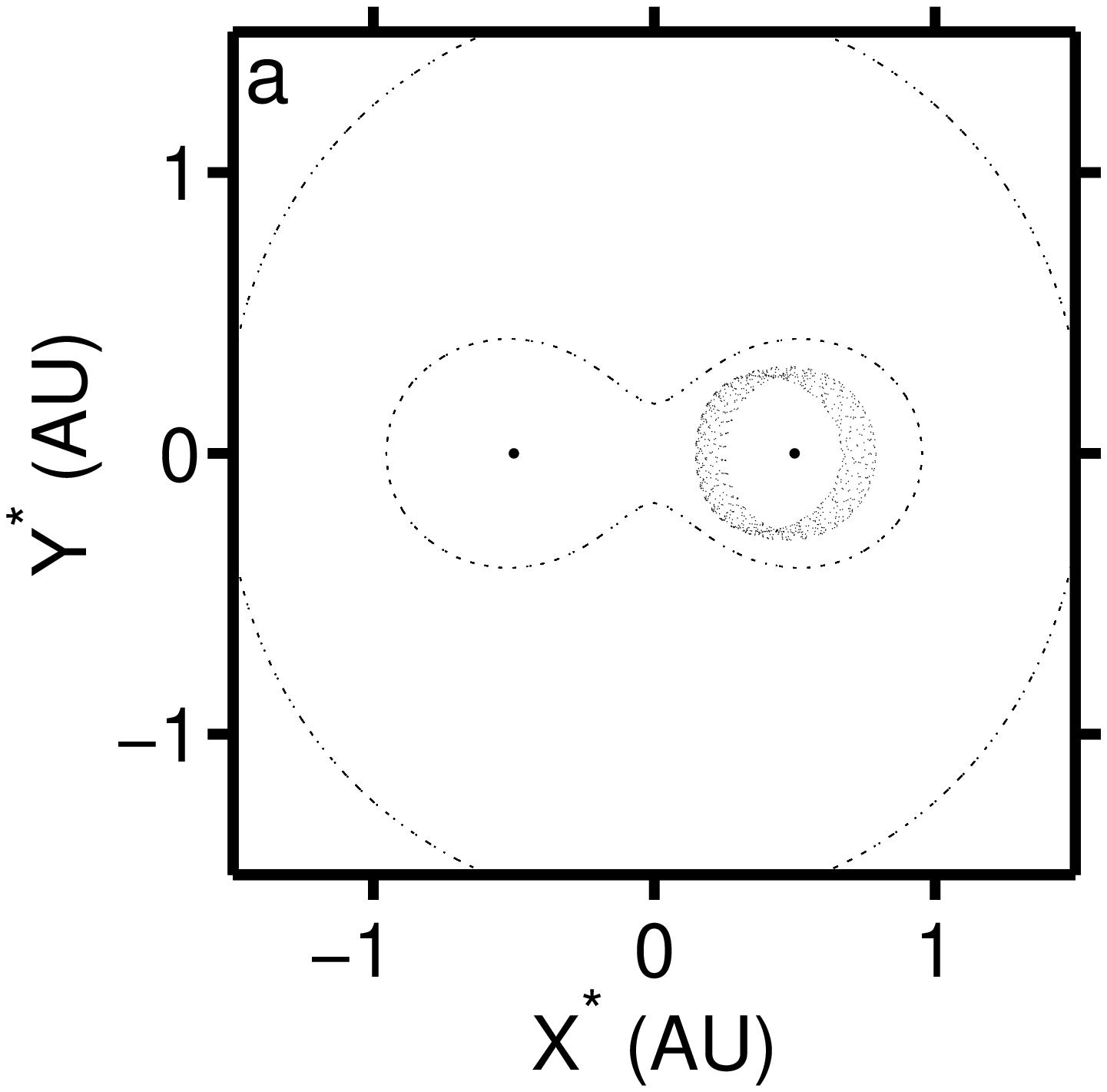,width=0.3\linewidth,height=0.3\linewidth}&
\epsfig{file=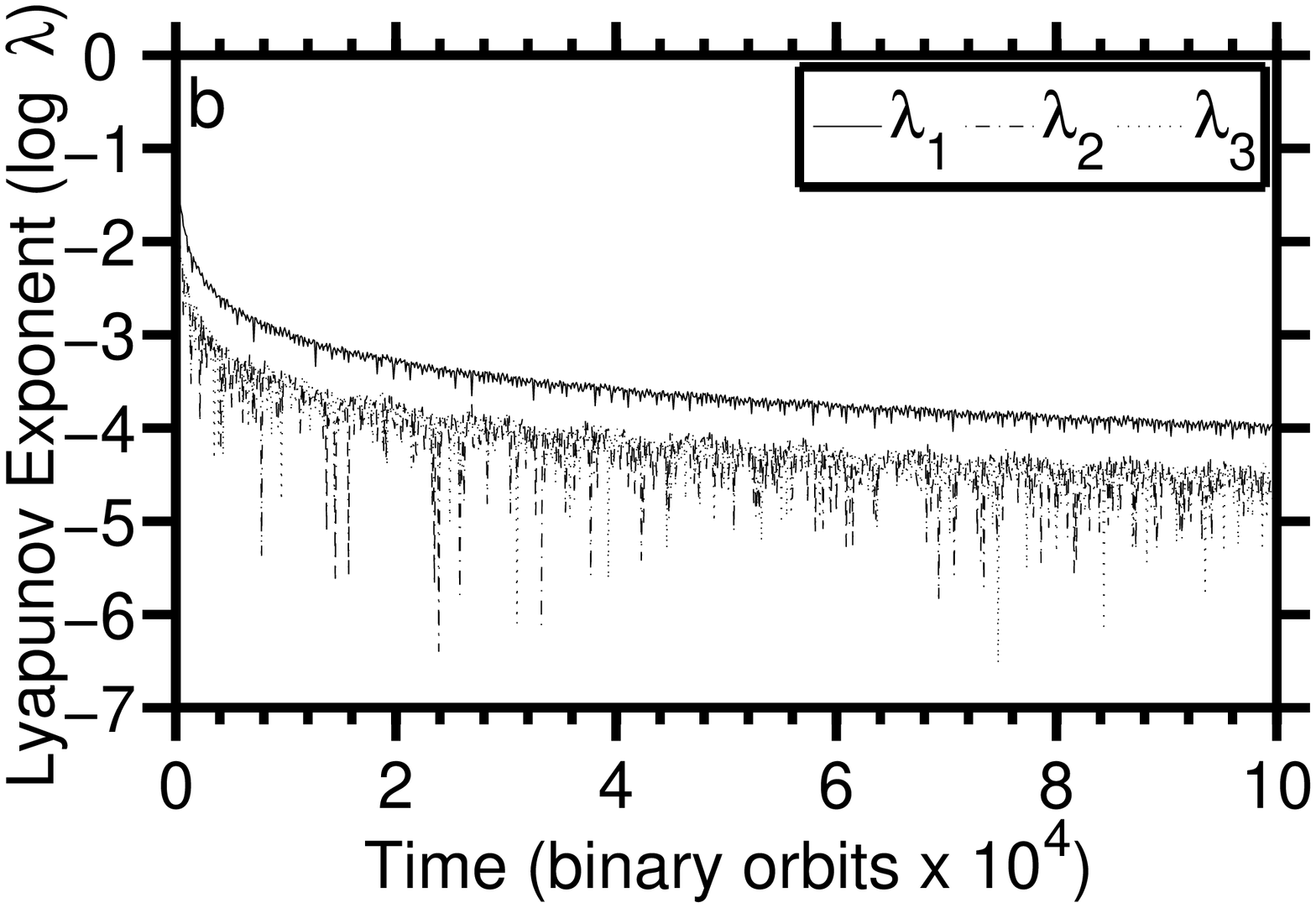,width=0.3\linewidth,height=0.3\linewidth}&
\epsfig{file=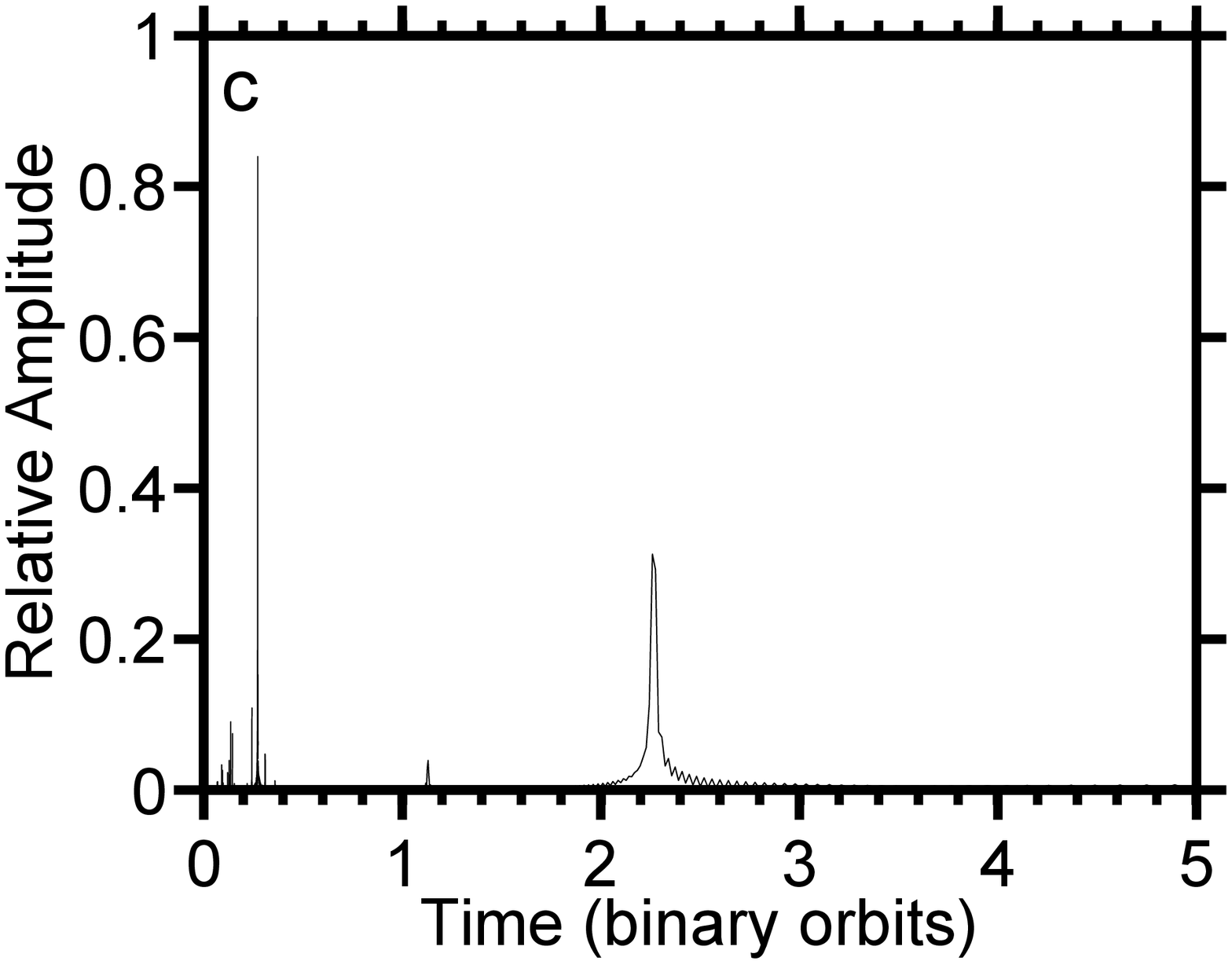,width=0.3\linewidth,height=0.3\linewidth}\\
\end{tabular}
\caption{Case study for the initial planetary distance ratio
$\rho_0 = 0.290$ with the planetary orbit in the synodic coordinate system, Lyapunov spectrum, 
and power spectrum.  The stellar mass ratio is $\mu = 0.5$.}
\label{fig:mu.5rho.29}
\end{figure*}

\begin{figure*} [h]
\centering
\begin{tabular} {ccc}
\epsfig{file=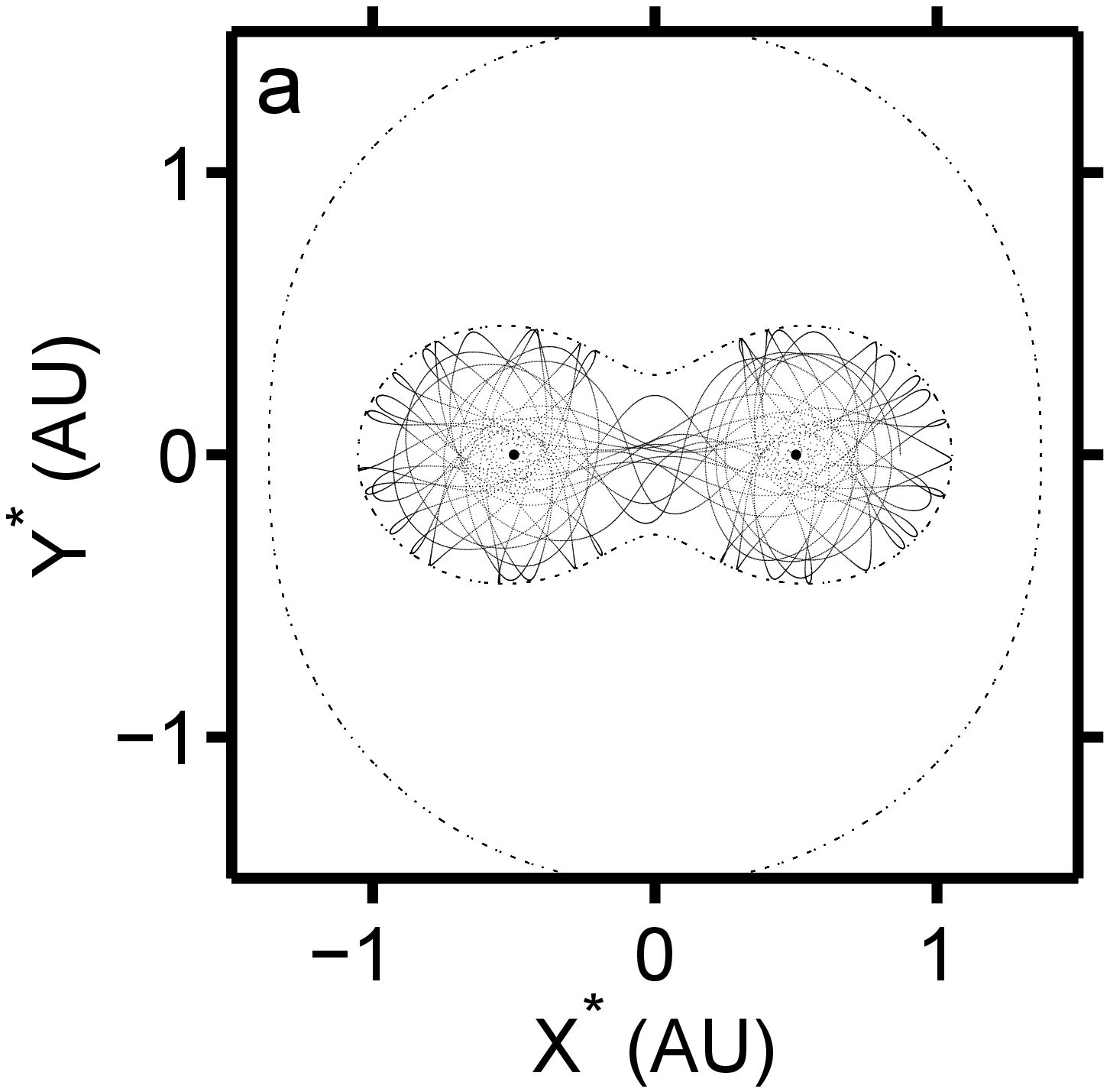,width=0.3\linewidth,height=0.3\linewidth}&
\epsfig{file=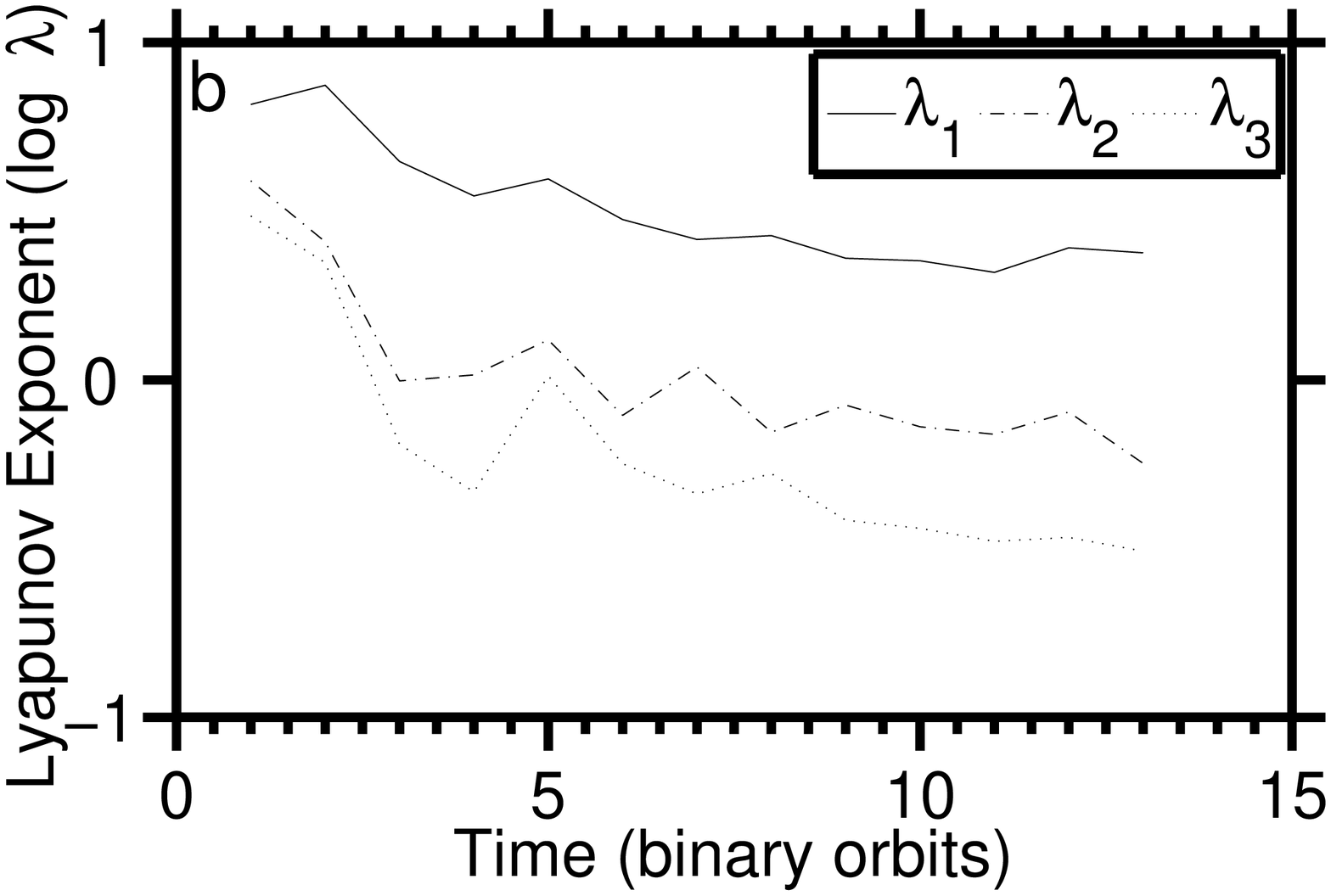,width=0.3\linewidth,height=0.3\linewidth}&
\epsfig{file=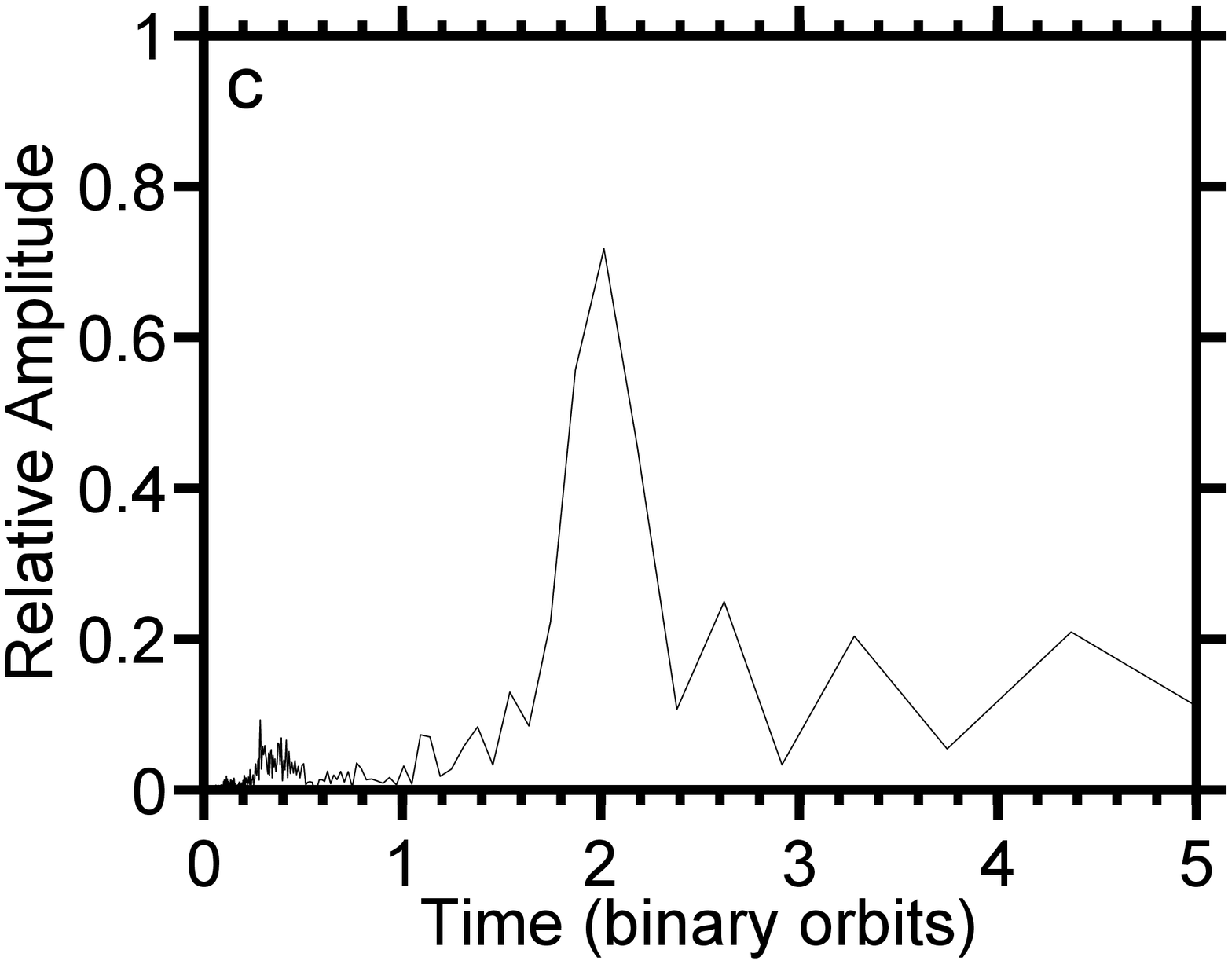,width=0.3\linewidth,height=0.3\linewidth}\\
\end{tabular}
\caption{Case study for the initial planetary distance ratio
$\rho_0 = 0.370$ with the planetary orbit in the synodic coordinate system, Lyapunov spectrum, 
and power spectrum.  The stellar mass ratio is $\mu = 0.5$.}
\label{fig:mu.5rho.37}
\end{figure*}

\begin{figure*} [h]
\centering
\begin{tabular} {ccc}
\epsfig{file=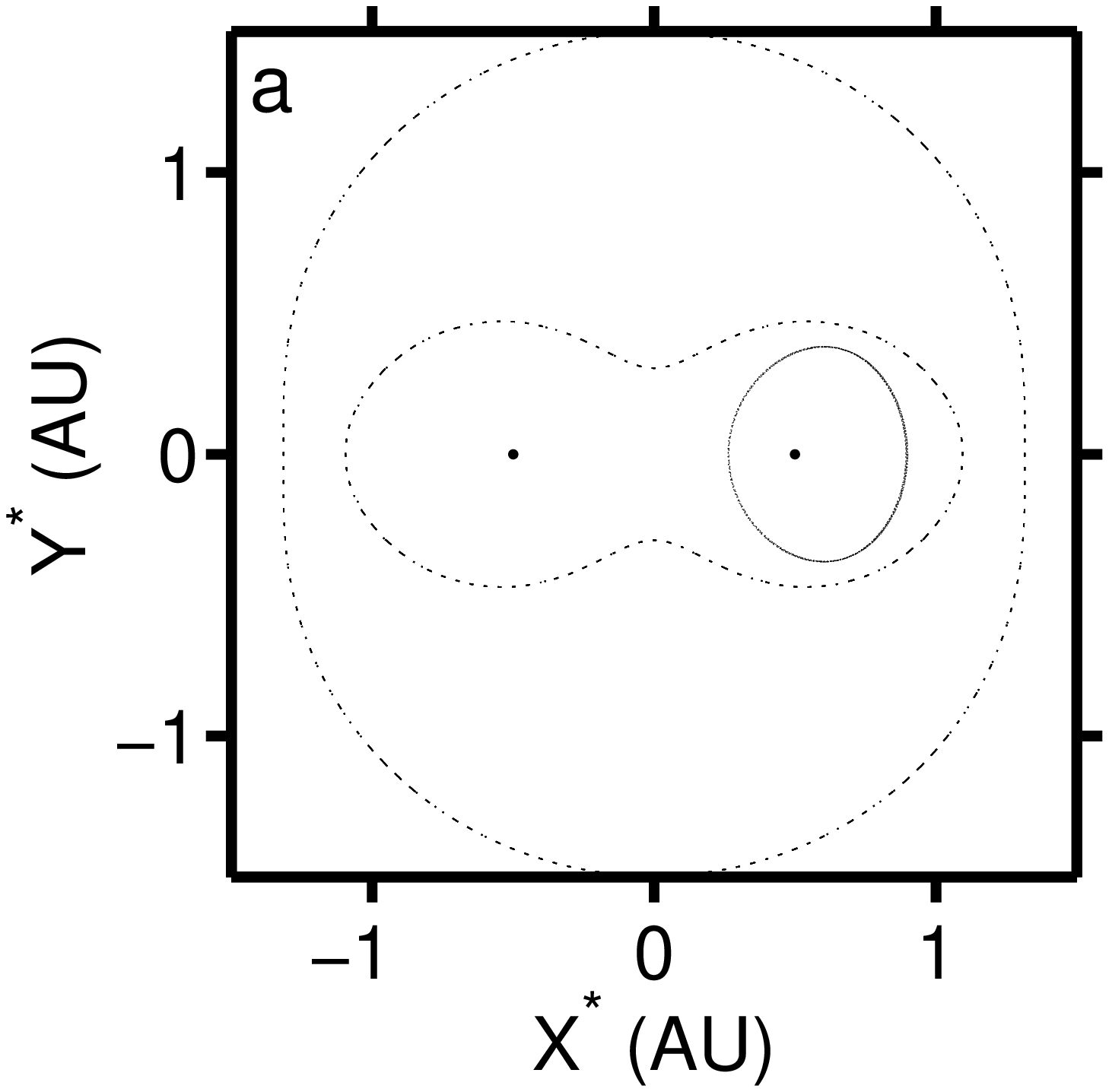,width=0.3\linewidth,height=0.3\linewidth}&
\epsfig{file=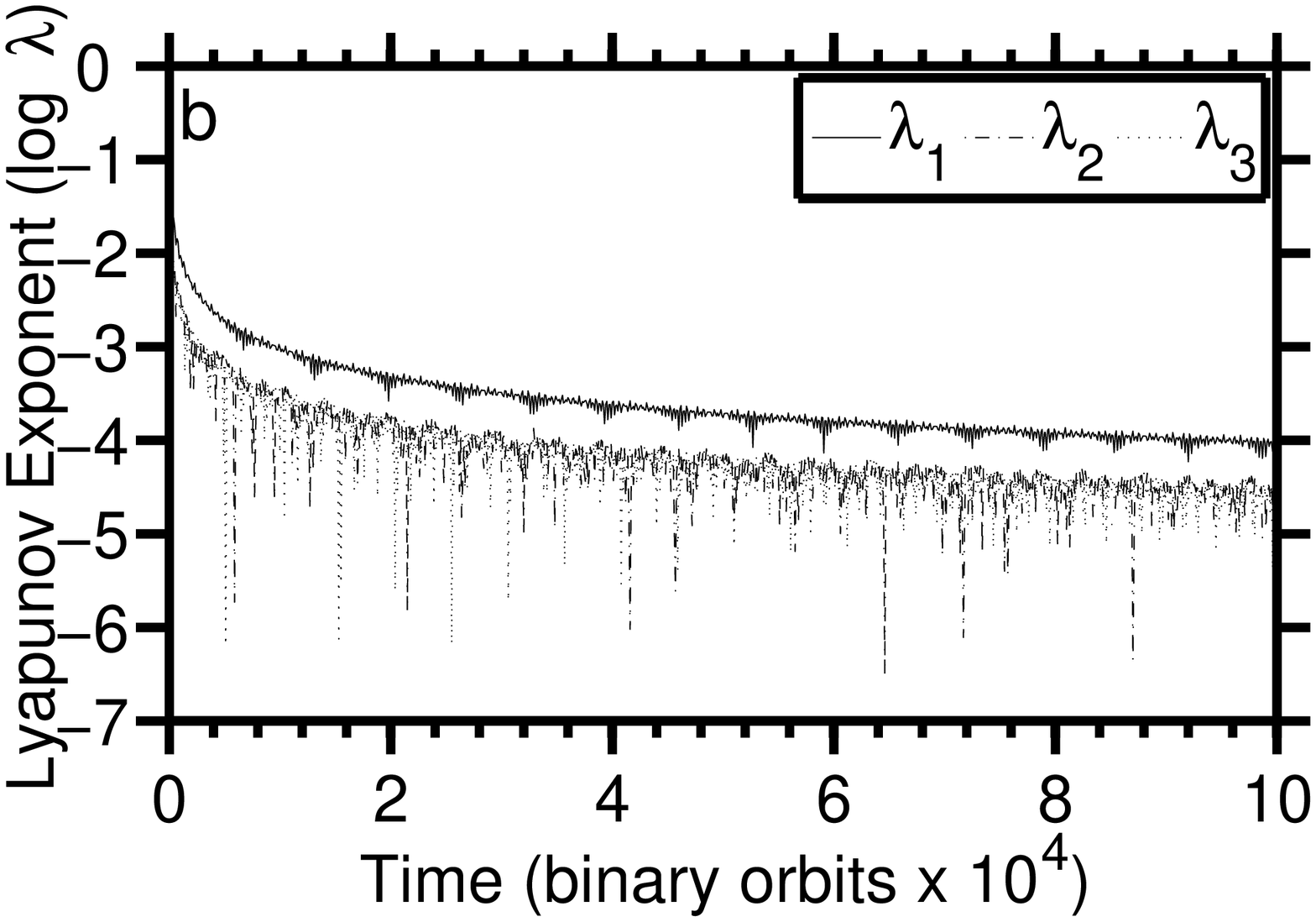,width=0.3\linewidth,height=0.3\linewidth}&
\epsfig{file=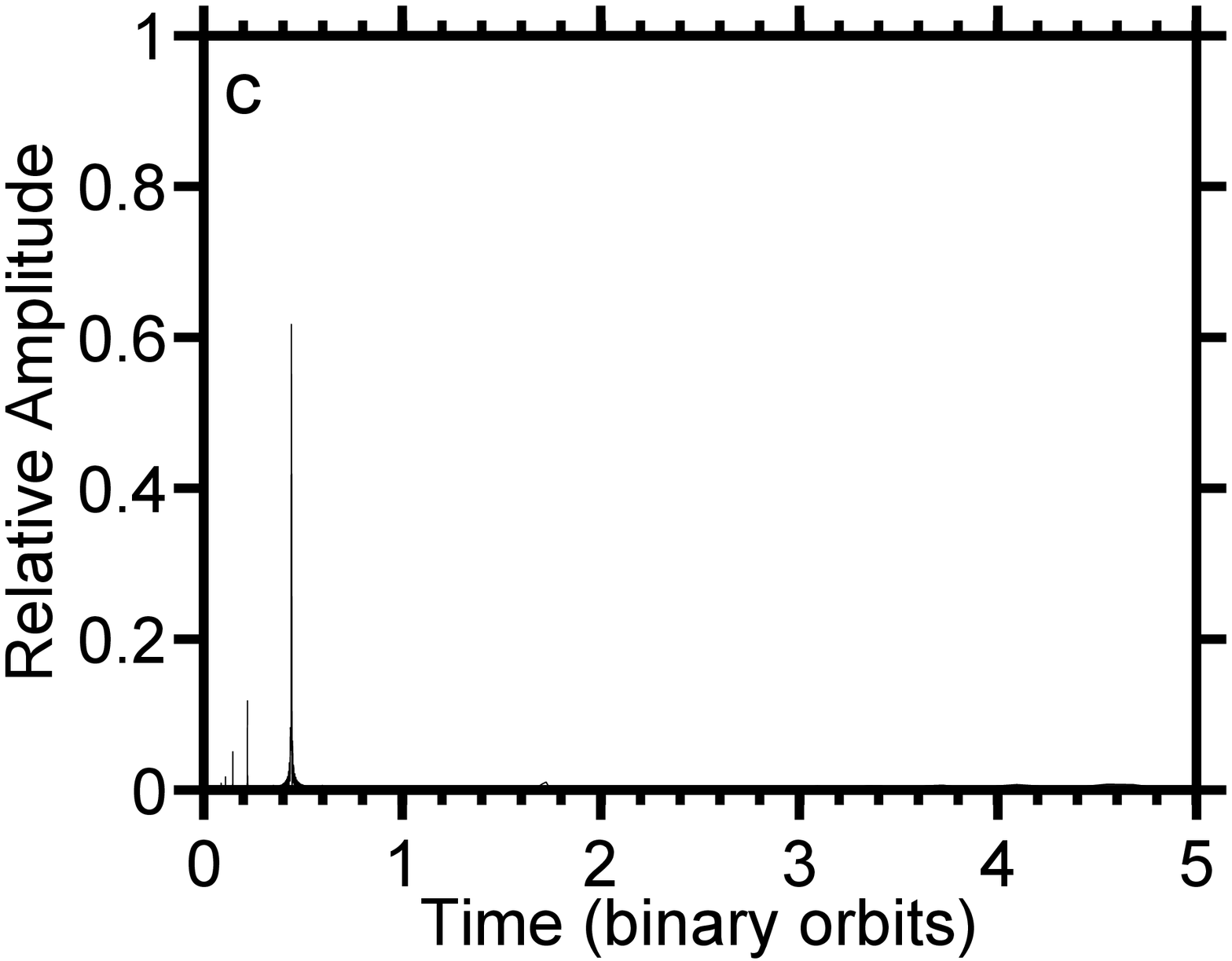,width=0.3\linewidth,height=0.3\linewidth}\\
\end{tabular}
\caption{Case study for the initial planetary distance ratio
$\rho_0 = 0.400$ with the planetary orbit in the synodic coordinate system, Lyapunov spectrum, 
and power spectrum.  The stellar mass ratio is $\mu = 0.5$.}
\label{fig:mu.5rho.40}
\end{figure*}

We display runs for selected initial conditions (i.e., starting distances $\rho_0$ 
of the planet from the stellar primary) corresponding to $\mu = 0.3$
in Figs.~\ref{fig:mu.3rho.355} to \ref{fig:mu.3rho.595} and to $\mu = 0.5$ in
Figs.~\ref{fig:mu.5rho.29} to \ref{fig:mu.5rho.40}.  In Figs.~\ref{fig:mu.3rho.355}
to \ref{fig:mu.5rho.40}, we first present the planetary orbits in the synodic coordinate 
system (X*, Y*).  Secondly in each figure, the Lyapunov spectra of the simulations
are shown using a logarithmic scale for the Y*-axis.  Lastly in each figure, the time
series power spectra of the simulations are shown with their normalized amplitudes.
The power spectra were obtained through the usage of a FFT subroutine in
Matlab$\textsuperscript{\textregistered}$ that uses the X*-component separation distance
as a function of time and converts the output frequencies to periods.
The selected starting distance ratios $\rho_0$ of the planet are 0.355, 0.474, 0.595
for $\mu = 0.3$ and 0.290, 0.370, 0.400 for $\mu = 0.5$.  Note that
$\rho_0$ indicates the relative initial distance of the planet given as $R_0/D$, with
$D = 1$~AU as the distance between the two stars and $R_0$ as the initial distance of
the planet from its host star, the primary star of the binary system.

Using the method of Lyapunov exponents, we are able to verify and extend the methods
described by \cite{ebe08} and \cite{ebe10}.  Absolute orbital stability can
be more rigorously shown through the Lyapunov exponent method.  It occurs when
$\rho_0 < \rho_0^{\left(1\right)}$, where $\rho_0^{\left(1\right)}$ represents the point
at which the initial condition results in a ZVC that opens at L1.  For larger values of
$\rho_0$, stability is not guaranteed due the behaviour discussed by \cite{ebe08}; 
this result is also consistent with our Lyapunov exponent criterion for orbital 
stability. 

\begin{table*}
\caption{Maximum Lyapunov exponent study for the models of $\mu = 0.5$ at different 
time intervals.}
\centering
\vspace{0.05in}
\vspace{0.05in}
\begin{tabular}{l c c c c c}
\hline
\hline
\noalign{\vspace{0.03in}}
$\rho_0$ & $\rm MLE\left({\rm 10^2}\right)$ & $\rm MLE\left({\rm 10^3}\right)$ 
& $\rm MLE\left({\rm 10^4}\right)$ & $e_{\rm median}$ & \rm ZVC  \\
\noalign{\vspace{0.03in}}
\hline
\noalign{\vspace{0.03in}}
 0.25  & 8.7189E$-2$\p* &  1.1353E$-2$\p* &  1.0009E$-3$\p* & 0.20  & ...        \\
 0.29  & 1.0647E$-1$\p* &  9.5469E$-3$\p* &  9.9141E$-4$\p* & 0.45  & L1         \\
 0.30  & 7.6275E$+0$*   &   ...           &   ...           & 0.67  & L1         \\
 0.35  & 5.2937E$+0$*   &   ...           &   ...           & 0.82  & L1         \\
 0.37  & 6.1946E$+0$*   &   ...           &   ...           & 0.98  & L1         \\
 0.40  & 1.0675E$-1$\p* &  9.6613E$-3$\p* &  8.8900E$-4$\p* & 0.71  & L1         \\
 0.43  & 2.8614E$-1$\p* &  2.9420E$-1$*   &   ...           & 0.87  & L1         \\
 0.50  & 7.9684E$-1$*   &   ...           &   ...           & 0.089 & L1, L2, L3 \\ 
\noalign{\vspace{0.05in}} \hline
\end{tabular}
\vspace{0.05in}
\begin{list}{}{}
\item[]Note: Elements without data indicate simulations that ended due to the
energy criterion, thus representing planetary catastrophes.  Elements with an
asterisk (*) indicate that the simulation ended before the allotted time.
Also, $e_{\rm median}$ represents the median of the eccentricity distribution
obtained for the curvature of the planetary orbits in the synodic coordinate system
(see Paper~II).  The last column indicates the Lagrange point(s) where the
zero-velocity contour (ZVC) is open (see Paper~I).
\end{list}
\label{table3}
\end{table*}

In this paper, the main criterion for orbital stability is the Lyapunov exponent
criterion, which is based on the maximum Lyapunov exponent.  From a 
theoretical point of view, an orbit is stable when all Lyapunov exponents are 
exactly zero.  Obviously, this `perfect' criterion for orbital stability will
be very difficult to reach numerically because it would require an infinite 
simulation time.  Based on our finite simulation times, we obtain 3 positive and 
3 negative Lyapunov exponents, and their sum becomes close to zero within the
limits of our numerical simulations.  Hence, in order to determine orbital 
stability numerically, we must look at the values of the three positive 
Lyapunov exponents at the beginning and at the end of our simulations.  By 
comparing these values, we determine whether the exponents decrease in
time, and if so what is the rate of their decrease, or whether they stay
approximately constant in time.  

Using this information, we are able to identify the maximum Lyapunov exponent
and adopt it as our primary indicator of orbital stability.  We classify an
orbit as stable if the initial value of the maximum Lyapunov exponent is below
a certain threshold and if it decreases at a `reasonable rate' (see below); 
otherwise, the orbit is classified as unstable.  In our plots of Lyapunov 
spectra (see the second panels in Figs.~\ref{fig:mu.3rho.355} to 
\ref{fig:mu.5rho.40}), we present all three positive Lyapunov exponents
to show their values and study how they change in time.

In addition to the Lyapunov exponent criterion for orbital stability, we also 
use the so-called orbital energy criterion, which requires that the kinetic 
energy should not exceed twice the value of the potential energy.  This is evaluated 
in our numerical simulations during each time step.  A failure of this criterion
would imply a break in conservation of the Jacobi constant as detailed by 
\cite{sze67} and shown numerically in Table~\ref{table1}.  It should also be
noted that the cases presented in Table \ref{table1} are also presented in
Figs.~\ref{fig:mu.3rho.355} to \ref{fig:mu.3rho.595}.
There are two important points associated with this criterion.
First, our systems are Hamiltonian, which means that their energy must be 
conserved.  On the other hand, the fact that the energy conservation of the 
planet may be broken because we are neglecting the effects of the third mass 
on the larger masses has already been discussed by \cite{sze67}.  Second,
even if the energy is only approximately conserved in our numerical simulations, 
we still request that the Jacobi constant remains constant, which is a required 
stability condition for CR3BP.  To be consistent with this criterion (see also 
Paper I), we stop our numerical simulation once changes (even small) in
the Jacobi constant occur; such cases are depicted in Tables~\ref{table2} and
\ref{table3}.   

Alike in the previous paper by \cite{ebe08}, we are able to identify the same 
regions of orbital stability, instability, and domains of quasi-periodic orbital 
stability.  A key difference, however, is that the orbit diagrams depicted in
Figs.~\ref{fig:mu.3rho.355} to \ref{fig:mu.5rho.40} have been simulated for $10^5$ 
years, whereas the corresponding figures in the previous paper have been simulated 
for only $10^3$ years.  The power spectra are shown to indicate how we determined 
the region of quasi-periodic orbital stability including the correct magnitude.  

We begin by classifying Figs. \ref{fig:mu.3rho.355}, \ref{fig:mu.3rho.474},
\ref{fig:mu.5rho.29}, and \ref{fig:mu.5rho.40} as stable candidate 
configurations.  Three of the cases 
show similar behaviour in the maximum Lyapunov exponent as shown in Figs.~\ref{fig:mu.3rho.474}b,
\ref{fig:mu.5rho.29}b, and \ref{fig:mu.5rho.40}b.  These Lyapunov spectra have a common trend 
by starting at a maximum 
value on the order of $10^{-1}$ and converging, albeit slowly, to smaller orders of ten.  
Figure~\ref{fig:mu.3rho.474} shows a somewhat different behaviour compared to the other cases in 
its class of stable candidates.  This case establishes a quasi-periodic orbit, which illustrates 
a $\left(3:1\right)$ resonance in the power spectrum and reveals the same trend of 
stability in the Lyapunov exponent spectrum.  Figure~\ref{fig:mu.3rho.355} shows an additional
variance in behaviour due to the elevated nature of the maximum Lyapunov exponent.
This shows that a limited amount of chaos exists in the system while remaining stable
for the full simulation time.  This case also demonstrates a $\left(4:1\right)$ resonance
in the power spectrum.

In contrast, Fig.~\ref{fig:mu.3rho.595} demonstrates a case of instability.  
The Lyapunov spectrum shows a different nature than that of the preceding cases.
In Fig.~\ref{fig:mu.3rho.595}b it begins at a maximum value greater than 1 and converges
to a value between $10^{-1}$ and 1.  By establishing the preceding cases as stable cases,
we can contrast the final values of the corresponding maximum Lyapunov exponents.  In the
unstable case of Fig.~\ref{fig:mu.3rho.595}b, it is two orders of magnitude greater than
in the stable cases of Figs.~\ref{fig:mu.3rho.474}b, \ref{fig:mu.5rho.29}b, and
\ref{fig:mu.5rho.40}b.  However, in comparison with Fig.~\ref{fig:mu.3rho.355}b we find
only a difference of one order of magnitude.

Figure~\ref{fig:mu.5rho.37} has been examined in a similar manner.  Having already classified
Fig.~\ref{fig:mu.5rho.29} and \ref{fig:mu.5rho.40} as stable configurations, we emphasize 
that they reveal similar trends in the maximum Lyapunov exponent as given by Fig.~\ref{fig:mu.3rho.474}.
However, Fig.~\ref{fig:mu.5rho.37} shows a 
different orbit diagram and is described by a maximum Lyapunov exponent that gives the  
outcome of instability.  This case conveys a much noisier power spectrum along with a maximum 
Lyapunov exponent on the order of 1 or greater.  Finer detail is illustrated in Table~\ref{table1}
and Table~\ref{table2}.  Particularly Table~\ref{table2} shows a boundary in the maximum
Lyapunov exponent where values near 0.1 and smaller for the first 100 years tend toward stability.

\begin{figure*}
\centering
\begin{tabular} {ll}
\epsfig{file=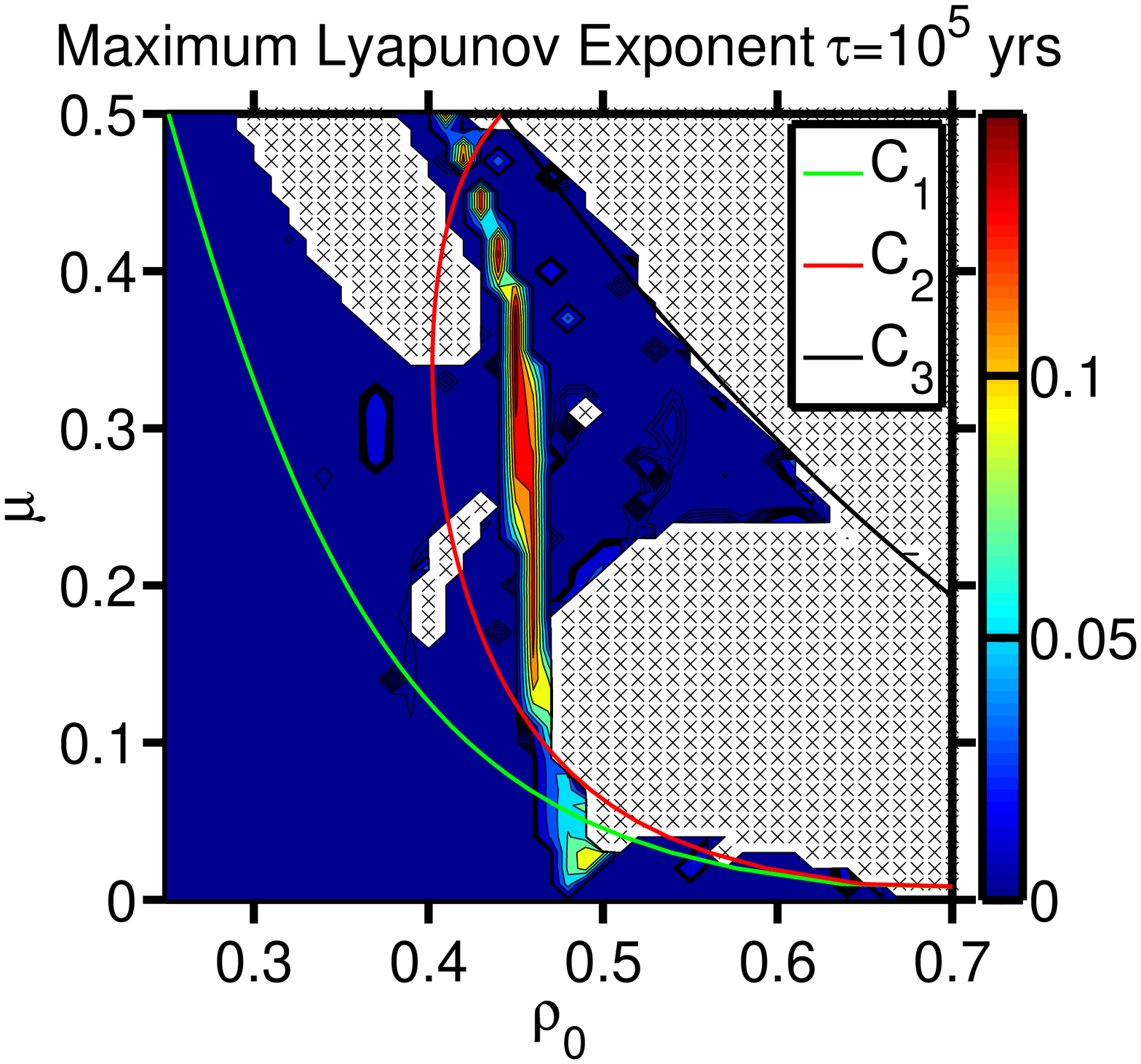,width=0.577\linewidth,height=0.375\linewidth}&
\epsfig{file=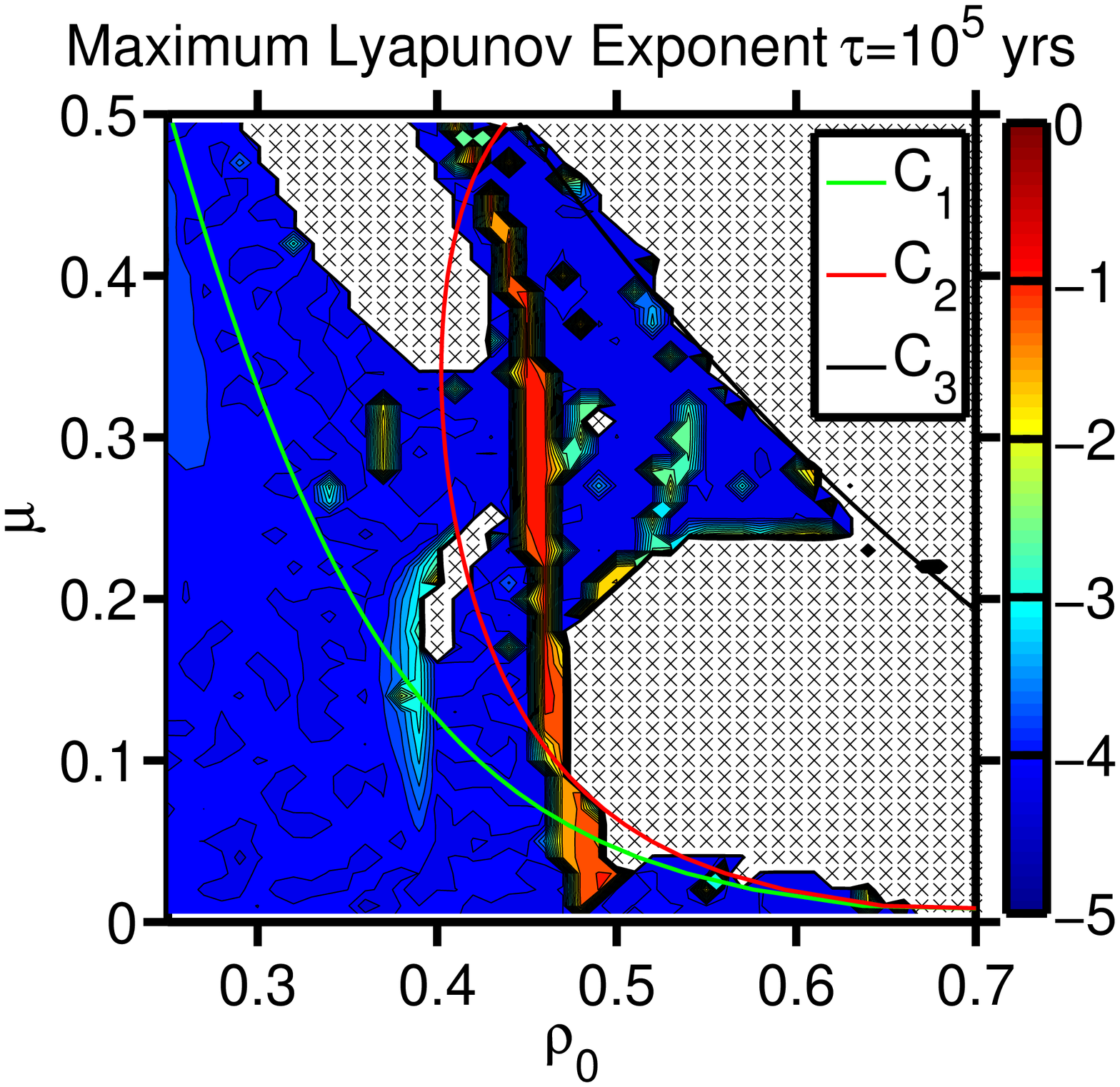,width=0.375\linewidth,height=0.375\linewidth}\\
\end{tabular}
\caption{We depict the maximum Lyapunov exponent (colour coded) for various mass ratios 
$\mu$ and initial conditions $\rho_0$ using a \emph{linear} scale (left) or a 
\emph{logarithmic} scale (right) where $-n$ corresponds to $10^{-n}$.  The crosses denote
cases where the simulation was terminated due to the planet being captured by one of the
stars or being ejected from the system. The green, red, and black curves represent the
initial conditions that cause the ZVC to open at L1, L2, and L3, respectively.}
\label{fig:contour}
\end{figure*}

Figure~\ref{fig:contour} conveys the bigger picture for the overall parameter space.
It represents contour plots of the maximum Lyapunov exponent with respect to the 
($\mu$, $\rho_{0}$) parameter space in linear and logarithmic scale.
The crosses depict initial conditions for runs that prematurely ended due to the orbital energy criterion.  
This demonstrates where the regions of instability occur as reflected by the respective colour coded scale 
of the maximum Lyapunov exponent.  Comparing Fig.~\ref{fig:contour} (left) to the previous result by
\cite{ebe08}, it can be seen that the main regions of stability remain the same.  Other noteworthy
features include that the instability islands present near $\rho_0 = 0.4$ have grown as the 
simulation time has been increased as well as the existence of a plateau near $\rho_0 = 0.48$.
The colour scale in Fig.~\ref{fig:contour} (left) has a maximum colour of dark red at a maximum
Lyapunov exponent of 0.15.
Some regions appear to be coloured black; however, this does not correspond to the adopted
colour scale as it is caused by the finite contour line width due to the close proximity of
the lines.  Therefore, the plateau has a peak value of a maximum Lyapunov exponent near 0.15.
Considering the diagrams in Fig.~\ref{fig:mu.3rho.474}, we conclude that a region of
quasi-periodicity exists on this plateau.

Figure~\ref{fig:contour} (right), the contour plot in logarithmic scale, is shown to display
some of the finer details pertaining to structure inside the stability regions.  The blue-green
coloured regions demonstrate areas of possible stable chaos that are hidden in
Fig.~\ref{fig:contour} (left).  Some other smaller contours are also produced in the stability region,
but the colour coding in general shows little change, which is partially due to the chosen spacing
between contours.  The average difference in height between these levels is $-0.15$; on the
logarithmic scale, it indicates a change by a factor of $1.41$ in the maximum Lyapunov exponent.

Inspecting Fig.~\ref{fig:contour} also allows a comparison with previous results, particularly
results of Paper~I.  There it was shown that absolute orbital stability occurs when
$\rho_0 < \rho_0^{\left(1\right)}$, where $\rho_0^{\left(1\right)}$ represents the point at
which due to the initial condition the ZVC opens at L1.  This condition is represented by the
line of C$_1$ in Fig.~\ref{fig:contour}.  It is evident that the Lyapunov exponent
criterion is almost perfectly consistent with this condition.  Additionally, as discussed in
Paper~I and references therein, no stability can be expected if $\rho_0 > \rho_0^{\left(3\right)}$
because now the ZVC is open at L1, L2, and L3.  This finding is also almost perfectly reflected
through the behaviour of the maximum Lyapunov exponent as a orbital stability criterion.  Both
panels of Fig.~\ref{fig:contour} show no stability regions on the right side of C$_3$, which is
reflective of the opening of the ZVC at Lagrange points L1, L2, and L3.

\section{Conclusions}

We present a detailed case study of the CR3BP with analyses through the usage of the Lyapunov
exponents.  We are able to characterize stability limits based on the value of the maximum 
Lyapunov exponent at the end of each simulation.  Cases where the maximum Lyapunov exponent 
exceeds a value of 0.15 indicate that the planet will experience an event that causes the orbital 
velocity to decrease.  These events include intersecting a Lagrange point or the ZVC. After 
such an event, the planet will experience a series of near misses (or collisions) with one of the stars in the 
binary system leading to overall instability.  Chaos theory and the concept of Lyapunov time 
prevent us from predicting exactly when the planet will be ejected.  Using the Lyapunov time as 
a measure of the length of predictive time, we can show that this relationship is proportional 
to the inverse of the maximum Lyapunov exponent.  Using our critical value of 0.15, the unstable 
systems will lose predictability within $(0.15)^{-1} = 6\frac{2}{3}$ years or less.  After that 
time, it is unknown when the planet will be ejected as this could take many multiples 
of the Lyapunov time to occur.

The method also shows evidence of a region of quasi-periodicity.  This is a region where the 
maximum Lyapunov exponent remains near the critical value without exceeding that value.  This 
is shown for cases simulated for $10^5$ years.  Based on this time scale, we can conclude that 
our quasi-periodic plateau represents a region of stable or quasi-stable chaos.  This is due 
to the Lyapunov times being much less than the simulation run time.  The chaos is shown through 
the value of the maximum Lyapunov exponent, whereas the stability is shown through the motion
of the planet over longer time scales.  The lack of near miss events and encounters with regions of 
decreasing velocity prohibits the planet from becoming orbitally unstable.

Comparisons to previously established criteria for stability show that our results are consistent
with previously obtained stability limits \citep[e.g.,][]{hol99,mus05,cun07,ebe08,ebe10}.
In \cite{ebe08}, Paper~I, the onset of instability was related to the topology of the ZVC,
whereas in \cite{ebe10}, Paper~II, it was shown that the onset of orbital instability occurs when
the median of the effective eccentricity distribution exceeds unity.  Both results are consistent
with the findings of Paper~III, i.e., the inspection of the maximum Lyapunov exponent.
However, the latter offers the advantage to link the study of orbital instability for the
CR3BP (and any subsequent generalization, if available) to chaos theory, including the
evaluation of different types of chaos.

Although our results have been obtained for the special case of the CR3BP, we expect that 
it may also be possible to augment our findings to planets in generalized stellar
binary systems.  Desired generalizations should include studies of the
elliptical restricted 3-body problem (ER3BP) \citep[e.g.,][]{pil02,sze08}
as well as of planets on inclined orbits \citep{dvo07}, noting that especially 
the former have important applications to real existing systems in consideration of the 
identified stellar and planetary orbital parameters \citep{egg10}.

\begin{acknowledgements}

This work has been supported by the U.S. Department of Education under GAANN
Grant No. P200A090284 (B.~Q. and J.~E.), the Alexander von Humboldt Foundation
(Z.~E.~M.) and the SETI institute (M.~C.).

\end{acknowledgements}

\section*{Appendix A}

\subsection*{Basic concepts and definitions}

From stability analysis we can use a Jacobian matrix to show how a state vector $\textbf{x}$ 
evolves in time (see \citealt{tso92} for details).  The governing equations are
\[\dot{\textbf{x}} = \textbf{J} \cdot \textbf{x}\] 
\[\textbf{J} = \left(\begin{array}{ccc}
{\partial f_{1} \over \partial x_{1}} & \ldots & {\partial f_{n} \over \partial x_{1}} \\
\vdots & \ddots &\vdots \\
{\partial f_{1} \over \partial x_{n}} & \ldots & {\partial f_{n} \over \partial x_{n}}
\end{array}\right)\]
This differential equation has a general solution, which is
$\textbf{x}\left(t\right) = e^{\textbf{J}t}\:\textbf{x}\left(0\right)$.  Assuming $\textbf{J}$ has distinct eigenvalues, 
we can find a matrix $\textit{U}$ to diagonalize $\textbf{J}$ to form a diagonal matrix 
$\textit{D}$ such as
\[\textit{U}^{-1}\:\textbf{J}\:\textit{U} = \textit{D}\]
By rewriting the above equation we obtain the more useful form given as
\[\textbf{J} = \textit{U}\: \textit{D}\: \textit{U}^{-1}\]
Using the multiplication theorem, we can develop relations to obtain the eigenvalues of $\textbf{J}$
yielding
\[\det\:\textbf{J} = \left(\det\:\textit{U}\right)\left(\det\: \textit{D}\right)\left(\det\: 
\textit{U}^{-1}\right) = \det\:\textit{D}\]
\[=\lambda_{1}\lambda_{2}\cdots\lambda_{n}\]
Furthermore, we find
\[Tr\:\textbf{J} = Tr\: \textit{D} = \lambda_{1} + \lambda_{2} + \ldots + \lambda_{n}\]
This can be generalized from $\textbf{J}$ to a function $f\left(\textbf{J}\right)$ such as
\[\det \: f\left(\textbf{J}\right) = f\left(\lambda_{1}\right)f\left(\lambda_{2}\right)\cdots f\left(\lambda_{n}\right)\]
\[Tr\: f\left(\textbf{J}\right) = f\left(\lambda_{1}\right) + f\left(\lambda_{2}\right) + \cdots + f\left(\lambda_{n}\right)\]
From our general solution, we can assume $f\left(\textbf{J}\right) = e^{\textbf{J}t}$.
Therefore, we find
\[\det\: e^{\textbf{J}t} = e^{\left(\lambda_{1} + \lambda_{2} + \ldots + \lambda_{n}\right)t} = e^{\left(Tr\:\textbf{J}\right)t}\]
A volume of perturbations in phase space will be conserved if $\left|\det\: e^{\left(Tr\:
\textbf{J}\right)t}\right| = 1$ or $Tr\:\textbf{J}=0$ for Hamiltonian systems.  A dissipative system 
will have $\left|\det\: e^{\left(Tr\:\textbf{J}\right)t}\right|<1$ or $Tr\:\textbf{J}<0$.  
The eigenvalues $\lambda_{1},\lambda_{2},\cdots,\lambda_{n}$ are the Lyapunov exponents of 
the flow or the logarithms of eigenvalues of $\textbf{J}$.

\subsection*{Numerical determination of Lyapunov exponents}

From these equations we can define 6 dimensional tangent vectors $\textbf{x}_{i}$ and their
derivatives $\dot{\textbf{x}}_{i}$ where $i=1,\ldots,6$ as
\[\textbf{x}_{i} = \left\{x_{i}, y_{i}, z_{i}, u_{i}, v_{i}, w_{i}\right\}^T \]
\[\dot{\textbf{x}}_{i} = \left\{u_{i}, v_{i}, w_{i}, \dot{u}_{i}, \dot{v}_{i}, 
\dot{w}_{i}\right\}^T \]
Also we can define a Jacobian matrix, $\textbf{J}$, given as
\[\textbf{J} = \left(\begin{array}{cccccc}
	0 & 0 & 0& 1 & 0 & 0\\
	0 & 0 & 0& 0 & 1 & 0\\
	0 & 0 & 0& 0 & 0 & 1\\
	{\partial \dot{u} \over \partial x} & {\partial \dot{u} \over \partial y} & {\partial \dot{u} 
\over \partial z}& 0 & 2 & 0\\
	{\partial \dot{v} \over \partial x} & {\partial \dot{v} \over \partial y} & {\partial \dot{v} 
	\over \partial z}& -2 & 0 & 0\\
	{\partial \dot{w} \over \partial x} & {\partial \dot{w} \over \partial y} & {\partial \dot{w} 
	\over \partial z}& 0 & 0 & 0\\
\end{array}\right)\]

\subsection*{Wolf method}
Now we describe the scheme by which \cite{wol85} determine the Lyapunov spectrum 
and its application to the CR3BP.  The Wolf method follows a basic algorithm.  The first step 
is to initialize a state vector of 6 elements.  Then, the tangent vectors need to be initialized 
to some value.  We choose to have all tangent vectors to be unit vectors for simplicity.  
This means that the elements $\left\{x_{1},\:y_{2},\:z_{3},\:u_{4},\:v_{5},\:w_{6}\right\} = 1$ 
and all other elements will be equal to zero.

The next step consists of a loop that will use a standard 
integrator akin to the Runge-Kutta schemes to determine how the state and tangent vectors will 
change within a time step.  This will continue for an adequate number of steps so that the tangent 
vectors become oriented along the flow.  When this has been accomplished, it is necessary 
to perform a Gram-Schmidt Renormalization (GSR) to orthogonalize the tangent space.  Thereafter, 
we take the logarithm of the length of each tangent vector to obtain the Lyapunov exponents. 
We then continue the loop.

\subsection*{Procedure}

We use the first tangent vector, $\textbf{x}_{1}$, to define the basis for the GSR process.  
Thus, the first step consists in normalizing this vector.  With the vectors of
the new orthonormal set of tangent vectors, denoted by primes ($\prime$), we find
\[\begin{array}{l}
\textbf{x}_1^{'} = {\textbf{x}_1 \over \left\|\textbf{x}_{1}\right\|} \\
\textbf{x}_2^{'} = {\textbf{x}_2 - \left\langle\textbf{x}_2,\:\textbf{x}_1^{'} \right\rangle 
\textbf{x}_1^{'} \over \left\|\textbf{x}_2 - \left\langle\textbf{x}_2,\:\textbf{x}_1^{'} 
\right\rangle \textbf{x}_1^{'}\right\|} \\
\vdots \\
\textbf{x}_6^{'} = {\textbf{x}_6 - \left\langle\textbf{x}_6,\:\textbf{x}_5^{'} \right\rangle 
\textbf{x}_5^{'} - \ldots -\left\langle\textbf{x}_6,\:\textbf{x}_1^{'} \right\rangle 
\textbf{x}_1^{'} \over \left\|\textbf{x}_6 - \left\langle\textbf{x}_6,\:\textbf{x}_5^{'} 
\right\rangle \textbf{x}_5^{'} - \ldots -\left\langle\textbf{x}_6,\:\textbf{x}_1^{'} 
\right\rangle \textbf{x}_1^{'}\right\|} \\
\end{array} \]
From this new set of tangent vectors, the Lyapunov exponents will be calculated considering the 
lengths of each vector.  Therefore, we find
\[\lambda_i = {1\over \tau}\left[\lambda_{i-1} + \log\left\|\textbf{x}_i - \sum_{j=1}^{i}
{\left\langle \textbf{x}_i,\:\textbf{x}_{j-1}^{'}\right\rangle\textbf{x}_{j-1}^{'}}\right\| 
\right]\] $\ \rm where\:\lambda_{o}=\textbf{x}_{o}^{'}=0$.


\end{document}